\documentclass[10pt,aps,prc,twocolumn,showpacs,showkeys,nofootinbib]{revtex4-1}
\usepackage{graphicx}
\usepackage{amsmath,amsfonts,mathrsfs}
\newcommand{\nl}{\nonumber\\ }
\newcommand{\pd}{\partial}
\newcommand{\gD}{\mathcal{D}}
\newcommand{\vc}[1]{\vec{#1}}

\newcommand{\vU}{U}

\newcommand{\vD}{\Delta}

\newcommand{\Elab}{E_{\rm lab}}
\newcommand{\agev}{A\text{GeV}}
\newcommand{\rmd}{{\mathrm d}}
\newcommand{\rmid}{{\rm id}}

\newcommand{\Tfrz}{T_{\rm f.o.}}

\def\lsim{\mathrel{\rlap{
\lower4pt\hbox{\hskip-3pt$\sim$}}
    \raise1pt\hbox{$<$}}}     
\def\gsim{\mathrel{\rlap{
\lower4pt\hbox{\hskip-3pt$\sim$}}
    \raise1pt\hbox{$>$}}}     

\DeclareMathOperator{\Div}{div}

\usepackage{xcolor}

\begin{document}


\title{Hybrid model with viscous relativistic hydrodynamics: a role of constraints on the shear-stress tensor}

\author{A. S. Khvorostukhin}
\email{hvorost@theor.jinr.ru}
\affiliation{Joint Institute for Nuclear Research, RU-141980 Dubna, Russia}
\affiliation{Institute of Applied Physics, Moldova Academy of Science, MD-2028 Kishineu, Moldova}

\author{E. E. Kolomeitsev}
\email{kolomei@theor.jinr.ru}
\affiliation{Matej Bel University, SK-97401 Banska Bystrica, Slovakia}
\affiliation{Joint Institute for Nuclear Research, RU-141980 Dubna, Russia}

\author{V. D. Toneev}
\affiliation{Joint Institute for Nuclear Research, RU-141980 Dubna, Russia}

\begin{abstract}
We present the hybrid hadron string dynamic (HydHSD) model connecting the parton-hadron-string dynamic model (PHSD) and a hydrodynamic model taking into account shear viscosity within the Israel-Stewart approach. The numerical scheme, initialization, and particlization procedure are discussed in detail. The performance of the code is tested on the pion and proton rapidity and transverse mass distributions calculated
for Au+Au and Pb+Pb collision at AGS--SPS energies. The influence of the switch time from transport to hydro models, the viscous parameter, and freeze-out time are discussed. Since the applicability of the  Israel-Stewart hydrodynamics assumes the perturbative character of the viscous stress tensor, $\pi^{\mu\nu}$, which should not exceed the ideal energy-momentum tensor, $T_{\rm id}^{\mu\nu}$, hydrodynamical codes usually rescale the shear stress tensor if the inequality  $\|\pi^{\mu\nu}\|\ll \|T_{\rm id}^{\mu\nu}\|$ is not fulfilled in some sense.
There are several conditions used in the literature and we analyze in detail the influence of different conditions and values of the cut-off parameter on observables.
We show that the form of the corresponding condition plays an important role in the sensitivity of hydrodynamic calculations to the viscous parameter -- a ratio of the shear viscosity to the entropy density, $\eta/s$. It is shown that the constraints used in the vHLLE and MUSIC models give the same results for the observables. With these constraints, the rapidity distributions and transverse momentum spectra are most sensitive to a change of the $\eta/s$ ratio. We demonstrate that these constraints do not guarantee that each element of the  $\pi^{\mu\nu}$ tensor is smaller than the corresponding element $T_{\rm id}^{\mu\nu}$. As an alternative, a
strict condition is used. When applied it reduces the sensitivity of the proton and pion momentum distributions to the viscosity parameter. We performed global fits the rapidity and transverse mass distribution of pion and protons. It was also found that $\eta/s$ as a function of the collision energy monotonically increases from $\Elab=6\agev$ up to $\Elab=40\agev$ and saturates for higher SPS energies. We observe that it is difficult to reproduce simultaneously pion and proton rapidity distribution within our model with the present choice of the equation of state without a phase transition.
\end{abstract}

\maketitle


\section{Introduction}

Hydrodynamics is a powerful phenomenological tool having a variety of wonderful properties. It allows one to take easily into account collective effects and the equation of state (EoS) of studied matter which cannot be completely described by microscopic models. Application of hydrodynamics to the theoretical description of high-energy nuclear collisions has been started with Landau's original work \cite{La53}. The actual status and successful story of the hydrodynamics approach in ultra-relativistic heavy-ion collision theory is reflected in review articles \cite{KH03,HS13,GJS13,JH15,DKK16,FHS17}.

A problem of heavy ion collision modeling is that hydrodynamics applicability conditions are violated at the early and final stages of a nucleus-nucleus interaction. The main condition assumes that the mean free path of quasiparticles in a system has to be smaller than the system size. It is clear that this condition is not  satisfied at the beginning and the end of a collision when the medium is far from the local equilibrium.

One way to get around the mentioned problem is to construct a hybrid model. Within hybrid models, one of which we developed in~\cite{HYDHSD2015}, the initial conditions for hydrodynamic equations, i.e. space distributions of the energy density, charge density, and velocity field, are calculated using a kinetic model.

In Ref.~\cite{HYDHSD2015} we formulated a hybrid model, called HydHSD, connecting an initial state generated by the PHSD~1.0 code with ideal hydrodynamics at a later stage. More realistic calculations of heavy-ion collisions at relativistic energies need to take into account a non-zero viscosity of the QCD matter~\cite{Song-Heinz08a,Song-Heinz08b}. This paper aims to extend the HydHSD model~\cite{HYDHSD2015}  to include the effect of shear viscosity in the hydrodynamical stage. Viscosity will be included within the standard the Israel-Stewart approach~\cite{IS}. The applicability of the viscous hydrodynamics requires that the dissipative (viscous) effects give sufficiently small corrections to the equilibrium quantities~\cite{MNR2010}. In practice, this means that the viscous part of the energy-momentum tensor should not exceed the ideal part of the tensor. There are several criteria used in the literature to compare these two parts, e.g. in the vHLLE~\cite{KHB2013} and MUSIC~\cite{MUSIC} models. We will analyze the performance of these various criteria and study how they influence the final momenta distributions of main hadrons. For our conservative study, we use a hadronic equation of state and try to describe rapidity spectra and transverse momentum distributions of protons and pions produced in relativistic nuclear collisions in the range from AGS to SPS energies, $6\,\agev \le\Elab \le 160\,\agev$.

The article is organized as follows. We start with the description of the set of viscous hydrodynamic equations in Sec.~\ref{hydroeq_sec} and how it is solved numerically, see Sec.~\ref{numerical_sec}. Sec.~\ref{initial_sec} is devoted to obtaining the initial conditions. In Sec.~\ref{observable_sec}, we describe the particlization procedure used to obtain the particle momentum distributions in the final state. The hadron EoS is presented in Sec.~\ref{eos_sec}. In Sec.~\ref{param_depend} we discuss in detail how variations of shear viscosity, freeze-out temperature, and the constraints on the shear stress tensor affect the particle momentum distributions within our model. Special attention is paid to the constraints on the shear stress tensor, see Sec.~\ref{VHLLEMUSIC}. In Sec.~\ref{fitsection} we try to fit simultaneously pion and proton momentum distributions for five collision energies.
Technical details of our numerical algorithm are given in Appendices.

\section{The Model}\label{sec:model}

\subsection{Equations of viscous hydrodynamics}
\label{hydroeq_sec}
The system undergoing hydrodynamic evolution is described by the set of equations~\cite{hydroabout}
\begin{subequations}
\begin{align}
\partial_\mu T^{\mu\nu}&=0,
\label{hydrobase-T}\\
\partial_\mu J^\mu &=0,
\label{hydrobase-J}
\end{align}
\label{hydrobase}
\end{subequations}
including an energy-momentum tensor $T^{\mu\nu}$ and a baryon current $J^\mu$. The equations represent the conservation laws of the total energy, momentum, and baryon charge.
Here and below we will use the Cartesian coordinates. In the general case of a non-ideal fluid when dissipation processes are possible, the energy-momentum tensor and the baryon current can be cast in the form~\cite{hydroabout}
\begin{align}
T^{\mu\nu}&=T_{\rm id}^{\mu\nu} -\Pi\Delta^{\mu\nu}+\pi^{\mu\nu}, \quad
J^\mu=n\, u^\mu + V^\mu\,,
\label{Tmunu}\\
T_{\rm id}^{\mu\nu}&=\varepsilon\, u^\mu u^\nu-P\Delta^{\mu\nu}\,,
\label{T-ideal} \\
\Delta^{\mu\nu}&=g^{\mu\nu}-u^\mu u^\nu\,,
\label{Deltadef}
\end{align}
where $T_{\rm id}^{\mu\nu}$  is the ideal part of the energy-momentum tensor, $\varepsilon$, $n$, and $P$ are the energy density,  the baryon density, and the pressure in the local reference frame (LRF), respectively, and $g^{\mu\nu}={\rm diag}(1,-1,-1,-1)$ is the metric tensor. The full energy-momentum tensor contains additional terms: the bulk pressure $\Pi$ and the shear stress tensor $\pi^{\mu\nu}$, and the baryon current includes in general also a diffusion current $V^\mu$. The 4-velocity $u^\mu$ is defined here as an eigenvector of the full energy-density tensor $T^\mu_\lambda\,u^\lambda=\varepsilon\, u^\mu$ (the Landau definition). It is normalized as $u_\lambda u^\lambda=1$ and can be written as $u^\mu = \gamma(1,\vc{v})$ through the 3-velocity $\vc{v}$ and $\gamma=(1-v^2)^{-1/2}$. From this definition of the flow velocity it follows that $\pi^{\mu\nu}$ is a traceless symmetric tensor satisfying the orthogonality relations:
\begin{align}
\label{orthogonality}
u_\mu\pi^{\mu\nu}&=0,\quad \pi^{\mu\nu}=\pi^{\nu\mu},\quad \pi^\mu_\mu=0.
\end{align}
Equations (\ref{hydrobase}) have to be supplemented by the EoS $P=P(\varepsilon,n)$.

If one considers a perfect fluid and puts $\pi^{\mu\nu}=0$, $\Pi=0$, and $V^\mu=0$, then the system of equations becomes closed and can be solved for $T^{0\nu}_{\rm id}$ and $J^0$ taken as independent variables. In the viscous case, however, we need some additional equations for $\pi^{\mu\nu}$, $\Pi$,  and $V^\mu$ which become independent dynamical variables. Below, for simplicity, we neglect the heat flux, i.e., $V^\mu = 0$ is assumed. Note that in this case, the Landau and Eckart frames coincide.

Studies performed in~\cite{DMNR2012} show that there can be infinitely many choices for the explicit form and coefficients in the equations of motion for $\pi^{\mu\nu}$ and $\Pi$. In this work, we follow the original Israel-Stewart framework~\cite{IS}, in which all viscous terms of the second order in gradients are suppressed. Additional quantities are governed by the relaxation-type equations
\begin{align}
(u^\lambda\partial_\lambda)\,\Pi &=-\frac{\Pi+\zeta\theta}{\tau_\Pi},
\quad \theta= \partial_\lambda u^\lambda\,,
\nonumber\\
(u^\lambda\partial_\lambda)\,\pi^{\mu\nu}&=-\frac{\pi^{\mu\nu}-\eta W^{\mu\nu}}{\tau_\pi}\, ,
\label{ISeqs}\\
W^{\mu\nu} &= \Delta^{\mu\lambda}\partial_\lambda u^\nu
+ \Delta^{\nu\lambda}\partial_\lambda u^\mu-\frac23\,\Delta^{\mu\nu}\, \theta\,,
\nonumber
\end{align}
where $\tau_\Pi$ and  $\tau_\pi$ are the relaxation times for the bulk pressure and the shear stress tensor while $\zeta$ and $\eta$ are the bulk and shear viscosity, respectively.  For vanishing relaxation times, $\tau_\pi=\tau_\Pi=0$, Eqs.~(\ref{ISeqs})
lock viscous terms $\Pi$ and $\pi^{\mu\nu}$ to their first-order values $-\zeta\theta$ and $\eta W^{\mu\nu}$, respectively, so that replacing them in Eqs.~(\ref{hydrobase}) and (\ref{Tmunu}) we recover the well-known Navier-Stokes formulaes.

Finally, we quote the expression for the entropy. We need only zero component of the entropy 4-vector, $s^\mu$, which reads for a cell as
\begin{align}
\label{s0def}
s^0&=\Big(s-\frac{\tau_\pi}{4T\eta}\,\pi^{\mu\nu}\pi_{\mu\nu}
-\frac{\tau_\Pi}{2T\zeta}\,\Pi^2\Big)\,u^0
\end{align}
with the temperature $T=T(\varepsilon, n)$ given by the EoS. Then the total entropy is the sum of $s^0$ over all cells multiplied by the cell volume. The dissipative part in Eq.~(\ref{s0def}) can be larger than the first term, $s$, in some cells. This can happen due to numerical errors and because viscous hydrodynamics, being applied to heavy-ion collisions, works at the edge of its applicability range. To get rid of these artifacts,  we exclude cells with temperatures $T<50$\,MeV in calculations of the total entropy.

Below in this work, we will neglect the bulk viscosity and put $\zeta=0$ and $\Pi=0$.
Then the system of hydrodynamic equations is closed by the expressions for $\tau_\pi$ and $\eta$, which in principle have to be calculated consistently with the EoS. However, in our calculations we use simplified relations~\cite{MHHN2014}
\begin{align}
\eta = k_\eta\,s,\quad\tau_\pi = \frac{5\eta}{\varepsilon+P}\,,
\label{eta-tau-def}
\end{align}
where $k_\eta={\rm const}$ and the entropy density $s=s(\varepsilon, n)$ is given by the EoS. The coefficient $k_\eta$ will be varied to reach the best agreement with experimental data.

\subsection{Numerical scheme}\label{numerical_sec}

For the numerical implementation, we, first of all, have to specify independent variables in the equations of motion (\ref{hydrobase}) and (\ref{ISeqs}). In viscous hydrodynamical codes, one usually takes the $J^0$ and $T^{0\nu}$ components of the energy-momentum tensor. In this case, the reconstruction of the LRF quantities such as energy and baryon densities and the 3-velocity of the fluid cell becomes a complicated problem~\cite{MNR2010,Muronga07}. Instead, we will use the components of the ideal-fluid tensor $T^{0\nu}_{\rm id}$ as independent variables, which allows us to apply relations (\ref{recover}), (\ref{recover-v}), and (\ref{recover-v-dir}) without a problem.
Then evolution of $T^{0\nu}_{\rm id}$ is described by the equation
\begin{align}
\label{hydrodecomposition}
    \pd_\mu T_{\rmid}^{\mu\nu}&=-\pd_\mu\pi^{\mu\nu}\,,
\end{align}
which is just a rewriting of Eq.~(\ref{hydrobase-T}).

As follows from Eq.~(\ref{orthogonality}), only five components of the $\pi^{\mu\nu}$ tensor are independent.
The other can be reconstructed if the cell velocity is known. However, for some choices of this five-component set, the reconstructed components can contain a singularity if an element of the vector $\vc{v}$ vanishes~\cite{MNR2010}. We select $\pi^{yy}$, $\pi^{zz}$, $\pi^{xy}$, $\pi^{xz}$, and $\pi^{yz}$ as independent ones in our implementation of the algorithm. As one can see from Eq.~(\ref{recover-pi}), a singularity is absent for such a choice if $v<1$.

For a numerical realization equations~(\ref{hydrobase-J}), (\ref{hydrodecomposition}), and (\ref{ISeqs}) can be rewritten in the form
\begin{align}
\label{shastaform}
\partial_t\vc{U}&+\sum_i\partial_i(v_i\vec{U})=\vec{S}\,,
\end{align}
where the 10-dimensional vector for generalized densities is
$$
\vc{U}=(J^0,T_{\rm id}^{00}, T_{\rm id}^{0x}, T_{\rm id}^{0y}, T_{\rm id}^{0z}, \pi^{xy}, \pi^{xz}, \pi^{yz},\pi^{yy},\pi^{zz})
$$
and the corresponding source terms $\vec{S}$ is given in Appendix~\ref{app:3Dsplitshasta}, see Eqs.~(\ref{Scons}) and (\ref{Spi}). This set of equations is solved numerically employing the SHASTA (the SHarp and Smooth Transport Algorithm) algorithm~\cite{SHASTA,SHASTARischke}. First, we tried to implement SHASTA following the numerical scheme outlined in Section 4.2 of Ref.~\cite{MNR2010} extending it to $3+1$ dimensions. The corresponding formulae are collected in Appendix~\ref{app:shasta}. However, if one uses the single-pass method for the time evolution and makes the full-time step without more ado, one cannot achieve the quadratic accuracy in time. We check on the example of the Bjorken expansion model that it leads to the development of large numerical fluctuations in the energy density, the longitudinal velocity, and in the elements of the viscous stress tensor (see discussion in Appendix~\ref{app:3Dsplitshasta} and Fig.~\ref{fig:Bjorken-num-comp}.
As an improvement, we use Heun's method to reach the quadratic precision in time.\footnote{Recall that in the standard implementation of SHASTA, e.g., for solving ideal hydrodynamics in UrQMD, the mid-point rule is used to achieve the second-order accuracy in time.} Although the improved algorithm works well for the model problem, it lacks stability in the full $3+1$ calculations. To overcome the problem, we apply the 3D splitting method~\cite{SHASTA,Rischke-rev} in combination with Heun's method for the time propagation.
Heun's method, which is also used in MUSIC~\cite{MUSIC}, is more appropriate for solving viscous hydrodynamics since it allows to use first-order approximations for the time derivatives in the source term, whereas the mid-point rule needs approximations of the second-order for these derivatives, see, e.g., \cite{KHB2013}.

In some cells the relaxation time $\tau_\pi$ given by Eq.~(\ref{eta-tau-def}) may become smaller than the calculation time step. Then, following the idea from Section 3.2 of Ref.~\cite{KHB2013}, we evolve $\pi^{\mu\nu}$ using the formal solution of Eq.~(\ref{ISeqs})\footnote{Equation~(\ref{pi_formal}) is applied before the antidiffusion step.}
\begin{align}
\label{pi_formal}
\pi^{\mu\nu}(t_{n+1})&=[\pi^{\mu\nu}(t_n)-\eta W^{\mu\nu}]e^{-\Delta t/(\gamma\tau_\pi)}+\eta W^{\mu\nu}.
\end{align}
This solution is applied for cells, where $\gamma\tau_\pi<\Delta t$ which guarantees the smallness of the exponent.

When one uses 3D splitting together with a two-step Runge-Kutta method, generally, there are two ways how to build the algorithm: (i) to obtain solutions with the second-order accuracy for the 1st, 2nd, and 3rd axis sequentially; (ii) to make the first step for all axes and then to make the second step also for all axes. As MUSIC and vHLLE codes, we follow the second way. It allows easily include the formal solution (\ref{pi_formal}) in the 3D splitting scheme (see Eq.~(\ref{pi-explit}) in Appendix \ref{app:3Dsplitshasta}).

As emphasized in Ref.~\cite{MNR2010}, it is important to verify the applicability of the hydrodynamic equations at each calculation step. This means that viscous effects are only corrections to the ideal fluid energy-momentum tensor, i.e., $|\pi^{\mu\nu}|<C|T_{\rm id}^{\mu\nu}|$, where $C$ is a constant of order, but smaller than, one. If these conditions are not satisfied, fluid dynamics may not give a reasonable description of the space-time evolution of the system and the numerical calculation can become unstable~\cite{Molnar09}.
Therefore, at each time step in each cell, we calculate the ratio

\begin{align}
q=q_{\rm S}=\max_{\mu,\nu}\frac{|\pi^{\mu\nu}|}{|T_{\rm id}^{\mu\nu}|},\quad \mbox{(S-cond.)}
\label{q-def-S}
\end{align}
and verify the fulfillment of the condition~\cite{MNR2010}
\begin{align}
q<C,
\label{piconstrain}
\end{align}
where $C$ is a predefined positive constant, $C<1$. If the opposite occurs we rescale the shear stress tensor as
\begin{align}
\pi^{\mu\nu}\to \pi^{\mu\nu}_{\rm corr}=\pi^{\mu\nu}\frac{C}{q}\,.
\label{rescale}
\end{align}
Such a rescaling procedure is frequently used in the literature~\cite{KHB2013,KHPB,VISHNU,MUSIC}; however, there are differences in how tensors $\pi^{\mu\nu}$ and $T_{\rm id}^{\mu\nu}$ are compared, i.e., the quantity $q$ is evaluated.
This aspect will be considered in detail in Section~\ref{VHLLEMUSIC}.
The condition (\ref{piconstrain}) evaluated with $q$ from Eq.~(\ref{q-def-S}) will be denoted as the strict (S-) condition, to distinguish it from other types of conditions used in other codes, which we discuss later in Section~\ref{VHLLEMUSIC}. By default, we assume $C=1$, unless the value of $C$ is specified explicitly.

Heavy-ion collisions at relativistic energies are believed to produce a deconfined, strongly coupled quark-gluon plasma (QGP)~\cite{Shuryak1,Shuryak2}. In the initial stages of the collision, during which the QGP is produced, the system is surely far from equilibrium and cannot be described by hydrodynamics. However, modeling based on near-ideal hydrodynamics strongly suggests that a hydrodynamic treatment becomes applicable rather quickly, e.g. for RHIC energies it happens on the time scale $\sim 0.1\,{\rm fm}/c$~\cite{Heinz2005}. Some aspects of the transition to the hydrodynamical regime in strongly coupled dynamics (like non-Abelian plasmas similar to the QGP) can be studied in theories, which possess dual gravitational descriptions, the best-known example being $N= 4$ supersymmetric Yang-Mills (SYM) theory~\cite{SYM1,SYM2}. Using this gauge/gravity duality, it is possible to study how quickly a far-from-equilibrium strongly-coupled non-Abelian plasma relaxes to a state, in which a hydrodynamic description is getting accurate, and to estimate energy and entropy of the formed system\footnote{It was noted~\cite{Romatschke2017,Attems2017} that the system created in high energy nuclear collisions reaches or at least comes close to equilibrium. In particular, it was realized that because of the expansion of the matter into the vacuum, the system would cool and thus freeze into a hadronic gas quickly. Thus, it became apparent that a fluid dynamic approximation to the system dynamics had to start early, on a time-scale of $\tau\sim 1$ fm/c or less.}.
Results obtained in \cite{Heller12,Wu11,Keegan16,Romatschke2017} can be interpreted as that the second-order hydrodynamics is applicable when\footnote{Let us also mention that one has to be careful with conclusions of \cite{Attems2017} since as is seen from Fig.~5 there, $P_T\neq0$ when $P_{\rm eq}=0$. It is not the case of usual matter.}
\begin{align}
\label{neareqcondition}
  |\pi^{\mu\nu}|<|T^{\mu\nu}_{\rm id}|.
\end{align}
There is a terminological disagreement in the literature on how one should characterize the initial state of the hydrodynamical evolution. The papers \cite{Heller12,Wu11,Keegan16,Romatschke2017} assiduously underline that neither local near-equilibrium nor near isotropy are required for hydrodynamics applicability. Indeed, the condition~(\ref{neareqcondition}) allows for a large anisotropy of the pressure and that the state of the system is far from equilibrium. On the other hand, the shear pressure satisfying inequality~(\ref{neareqcondition}) or weaker one is often considered in the literature as one giving sufficiently small corrections to the equilibrium quantities, see, e.g., \cite{MNR2010,KHB2013}. Such a statement is not completely clear, since viscous contributions are large at least when $|\pi^{\mu\nu}|\gtrsim\frac12|T^{\mu\nu}_{\rm id}|$.
To be specific in the further discussions we will speak about a close/near to equilibrium state having in mind that inequality~(\ref{neareqcondition}) is fulfilled.

\begin{figure}
\centering
\includegraphics[width=7.5cm]{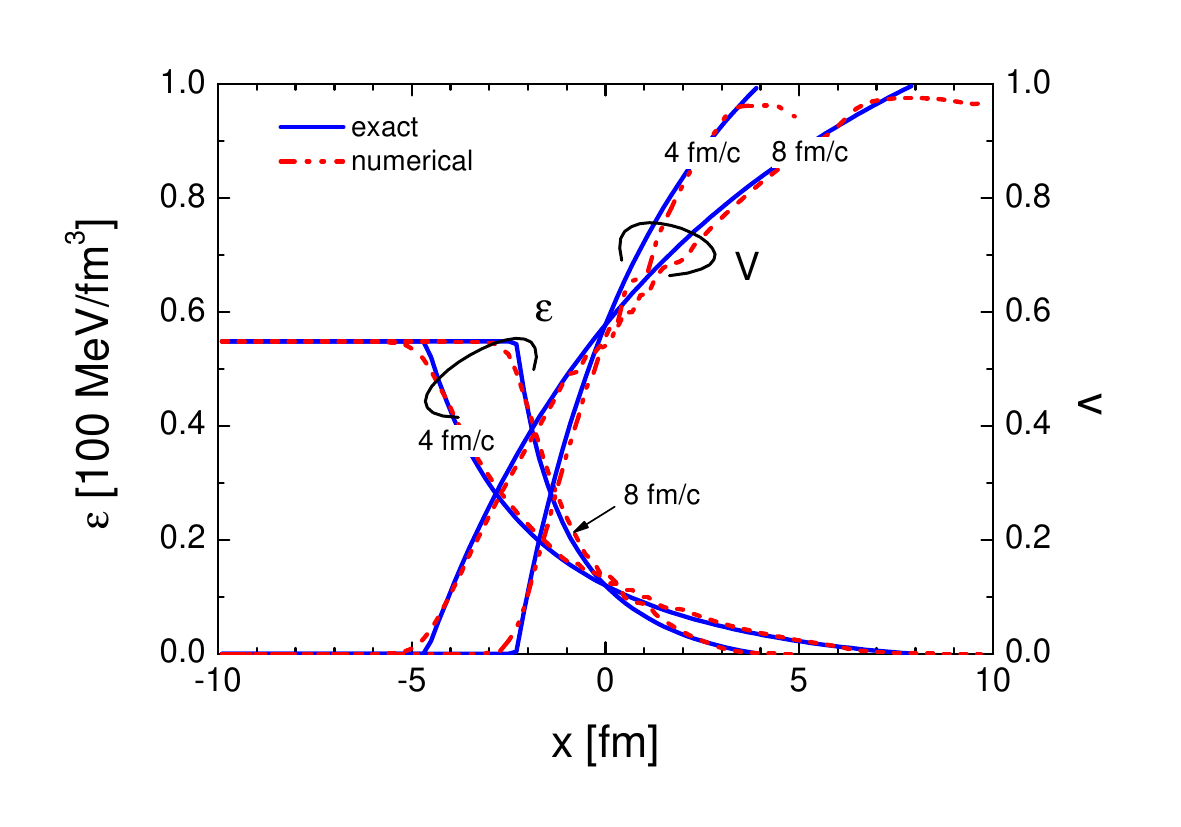}
\caption{The test of the hydrodynamical code on the solution of the (1+1)-dimensional Riemann problem. Solid lines show the analytic solutions~\cite{Skokov-Toneev07}. The energy density (left scale) and velocity (right scale) calculated by our algorithm at two moments of time (4\,{\rm fm}/$c$ and 8\,{\rm fm}/$c$ indicated by labels on the curves) are shown by solid lines.
} \label{analyttest}
\end{figure}
To verify our hydrodynamical code, we performed a test similar to that proposed in Ref.~\cite{MNR2010}, namely, we solved numerically the (1+1)-dimensional Riemann problem for two states with a constant pressure (or an energy density since $p=\varepsilon/3$) equal to $p_0$ on one side and to zero on the other side (vacuum) separated by a membrane located at $x = 0$. The evolution of energy density and velocity profiles is presented in Figs.~\ref{analyttest}. We used here $\Delta x = 0.2$\,fm and put very small shear viscosity, $\eta/s=0.01$ to simulate numerically a viscous free flow. We see a good agreement of numerical solutions with analytical ones.


We also checked our algorithm by 1D boost-invariant expansion \cite{ECHOQGP}, see Appendix~\ref{app:numerics}.

\begin{figure}
\centering
\includegraphics[width=6cm]{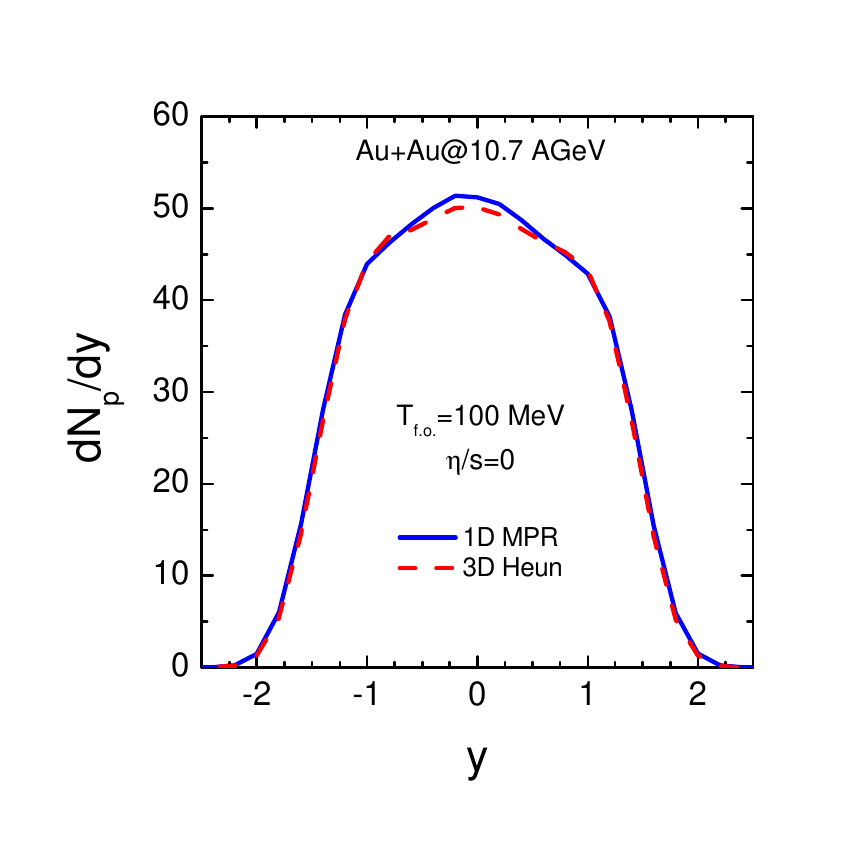}
\caption{The comparison of proton rapidity distributions for  Au+Au collisions at $\Elab=10.7\,\agev$ a the freeze-out temperature $\Tfrz=100$\,MeV calculated in ideal hydrodynamics using versions of SHASTA with the 1D mid-point rule (MPR) and with the 3D Heun (3D Heun) method for the time evolution.}\label{dNdy_E10_T100}
\end{figure}

Here as in our previous work~\cite{HYDHSD2015}, we solved equations of hydrodynamics by SHASTA with `phoenical' antidiffusion~\cite{SHASTARischke} and used the operator-splitting method to treat three-dimensional operators. Usually, one uses the mid-point rules (MPR) inside 1D propagation in applications of SHASTA to ideal hydrodynamics, i.e., three 1D steps each of which includes two Runge-Kutta steps. Our viscous code implements Heun's method (trapezoidal rule) and includes two sequential cycles over three axes. So, as an additional test of our code, we compare proton rapidity distributions for Au+Au collisions at $\Elab=10.7\,\agev$ calculated within ideal hydrodynamics applying the usual 1D MPR version and the new 3D Heun version of SHASTA. The corresponding results for the Au+Au collision at $\Elab=10.7\,\agev$ with the freeze-out temperature $\Tfrz=100$ MeV are shown in Fig.~\ref{dNdy_E10_T100}. One sees that the results are almost the same.
For our calculations, we take\footnote{Generally, one can use different mask coefficients, $A_{\rm ad}^{x,y,z}$, see Eq.~(\ref{Ad-mask}), for every direction, $x$, $y$, and $z$, but for simplicity we take one value for all axes.} $A_{\rm ad}=1$.


\subsection{Initialization of hydrodynamic evolution}\label{initial_sec}

The differential equations of hydrodynamics must be supplemented by appropriate initial conditions. As was noted in Section~\ref{numerical_sec}, in the initial stage a deconfined strongly coupled QGP phase may be formed.
In hybrid models, these conditions are usually deduced from results of calculations within some kinetic model, like the UrQMD model in the vHLLE code~\cite{KHB2013}. It would also allow for an event-by-event analysis of collisions. In our approach, we use the Hadron String Dynamics (HSD) model~\cite{HSD-1,HSD-2,HSD-3} which is very successful in the description of experimental data in the considered energy range.\footnote{Particularly we use version 1.0 of the Parton-Hadron String Dynamics model with the switched off partonic option.} To obtain relatively smooth initial distributions of the energy-momentum density and the baryon number, one can either perform averaging over many collision events or smear particles for a selected event in space with the help of a Gaussian distribution, for example~\cite{Oliin-Petersen15}. In our approach, we calculate the quantities
\begin{align}
T^{\mu\nu}_{\rm init} (\vec{r}\,) &=\overline{
\sum_a\frac{p_a^\mu\,p_a^\nu}{p_a^0} K(\vec{r}-\vec{r}_a)
}\,,
\nonumber\\
J^\mu_{\rm init}(\vec{r}\,) &=\overline{
\sum_a\frac{p_a^\mu}{p_a^0} K(\vec{r}-\vec{r}_a)
}\,,
\label{TJ-init}
\end{align}
where the bar stands for the event averaging and the sum runs over particles at the positions $\vec{r}_a$, $K(\vec{r}\,)$ is a smoothing function which in our case just averages over the volume element, $\Delta V$,
\begin{align}
K(\vec{r}\,)=\left\{\begin{array}{ll}
1/\Delta V &,\,\, \vec{r}\in\Delta V\\
0 &, \,\, \vec{r}\notin\Delta V
\end{array}
\right.
\,.
\end{align}
There are several methods of how to transit from $T_{\rm init}^{\mu\nu}$ and $J^\mu_{\rm init}$ to hydrodynamical quantities from~\cite{Oliin-Petersen15,GGHLO11}.
One is the procedure of an `ideal' initialization (IIS). One assumes that the structures of tensor $T_{\rm init}^{\mu\nu}$ and vector $J^\mu_{\rm init}$ are the same as for an ideal fluid, see Eqs.~(\ref{Tmunu}) and (\ref{T-ideal}). Then from quantities~(\ref{TJ-init}), one obtains the initial energy density, $\epsilon_{{\rm init}}$, and the baryon density, $n_{{\rm init}}$ in a fluid cell and the cell velocity, $\vec{v}$, with the help of relations (\ref{recover}), (\ref{recover-v}), and (\ref{recover-v-dir}). Such an approach is used, for instance, in the VHLLE code~\cite{KHB2013}. The initial entropy and other thermodynamical quantities are evaluated in each fluid cell using the equation of state, e.g., $s_{{\rm init}}=s(\epsilon_{{\rm init}}, n_{{\rm init}})$\,, cf. Eq. (\ref{s0def}) with $\pi^{\mu\nu}=0$ and $\Pi=0$. The initial total entropy and the baryon number of the system are finally calculated as
\begin{align}
S_{\rm init}=\sum_{\rm cell} \frac{s^0_{{\rm init}}}{\sqrt{1-\vec{v}^2}}\,,
\nonumber\\
N_{\rm init}=\sum_{\rm cell} \frac{n_{{\rm init}}}{\sqrt{1-\vec{v}^2}}\,.
\label{SN-init}
\end{align}
The advantage of this method is that it conserves the total energy, the total momentum, and the baryon number at the transition from HSD to the hydro regime. However, this procedure supports switching only to an ideal fluid, neglecting viscous corrections. Therefore, at the beginning of the hydrodynamical stage all components of the shear-stress tensor are initialized with zero values. That is found to be a useful approximation in the literature.

\begin{figure}
\includegraphics[width=5cm]{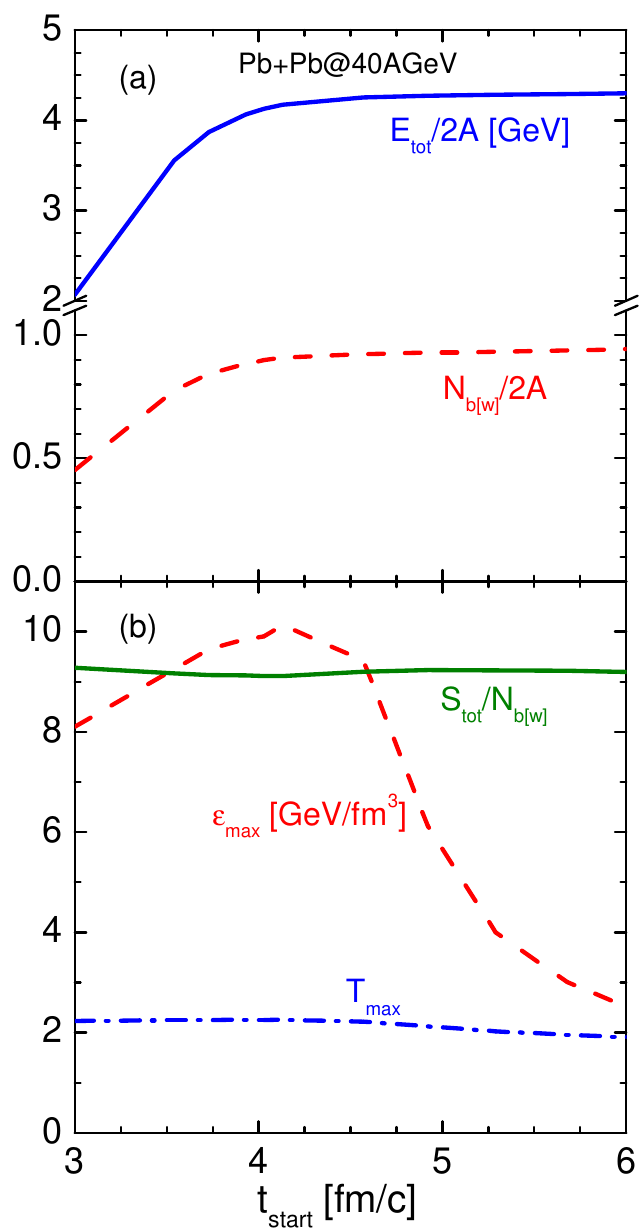}
\caption{Parameters of the initial stage of the hydrodynamic evolution as functions of the time, $t_{\rm start}$ of the transition from the kinetic (HSD) description to the hydrodynamics. Calculation are done for Pb$+$Pb ($A=208$) collisions at $\Elab=40\,\agev$ with the exact initialization procedure.  Panel (a) shows the total energy of stored in the fluid, $E_{\rm tot}$ and the number of wounded baryons, $N_{\rm b[w]}$. Panel (b) shows the entropy per baryon $S_{\rm tot}/N_{\rm b[w]}$, the maximum energy density, $\varepsilon_{\rm max}$, and the maximum temperature, $T_{\rm max}$.
}
\label{fig:t-start}
\end{figure}

The IIS method takes into account only four components of $T^{\mu\nu}$.
Another way to set up the energy-momentum tensor is to use all components calculated from a kinetic model with the help of Eq.~(\ref{TJ-init}). We will call it the `exact' initialization (EIS). To find initial energy density and velocity in a cell, one solves the eigenvalue problem $u_\mu T^{\mu\nu}=\varepsilon u^\mu$ which leads to the algebraic equation of fourth power (\ref{e-eq}) with coefficients~(\ref{e-eq-coeff}). The solution of this equation determines the velocity of the fluid (\ref{v-eq}). Then other hydrodynamical quantities follow with the help of Eqs.~(\ref{n-eq}), (\ref{piPi-eq}), and (\ref{V-eq}). Since we neglect the bulk pressure and the heat flow, the energy-momentum tensor is slightly not conserved when the `kinetic-to-hydro' transition is treated within the exact initialization.

A transition from a kinetic to a hydrodynamic regime occurs at an instant $t_{\rm start}$. We assume that at this moment the system is close to equilibrium and the ratio of the entropy to the baryon number $S(t)/N(t)$ ceases changing, see Fig.~1 in Ref.~\cite{HYDHSD2015}. To compute the time derivatives at the first step of the numerical solution, we need to know also the velocity and the shear stress tensor at the previous time step, $t_{-1}=t_{\rm start}-\Delta t$. For exact initialization, we obtain it from kinetic model by interpolation of $T^{\mu\nu}$ between $t_{\rm start}$ and $t_{\rm start}-\Delta t_{\rm kin}$. For ideal version of initialization procedure we put $\pi^{\mu\nu}(t_{-1})=0$ and $v_i(t_{-1})=0$.

Dependence of the parameters of the initial state in the hydrodynamics evolution on the transition time $t_{\rm start}$ is illustrated in Fig.~\ref{fig:t-start} for Pb$+$Pb collisions at $\Elab = 40$. The total energy and the number of participating (wounded) baryons shown in Fig.~\ref{fig:t-start}a saturates at times $t_{\rm star}\simeq 4\,{\rm fm}/c$, the entropy per baryon also stays almost constant for $t_{\rm start}>4\,{\rm fm}/c$. However, the energy density and temperature distributions can strongly depend on $t_{\rm start}$, for example, as shown in Fig.~\ref{fig:t-start}b the maximal values of the energy density decreases rapidly with an increase of $t_{\rm start}$ for $t_{\rm start}>4\,{\rm fm}/c$ because of the expansion of the system, the maximal temperature decreases also but much weaker.

Below we consider the following heavy-ion collisions:
Au$+$Au collisions for AGS energies at $\Elab = 6$ and 10.7\,$\agev$, and
Pb$+$Pb collisions for SPS energies at $\Elab = 40$, 80, and 158\,$\agev$.
All calculations are performed for the impact parameter $b=1$\,fm.

\subsection{Particlization procedure and observables}
\label{observable_sec}

To convert fluids to particles, we realized a particlization procedure  according to the Cooper-Frye formula~\cite{HYDHSD2015}:
\begin{align}
\label{CooperFrye}
E\frac{\rmd^3N_a}{\rmd p^3}&=\frac{g_a}{(2\pi)^3}\int\rmd\sigma_\mu  p^\mu f_a(x,p)\,, \end{align}
where $p^\mu=(E,\vec{p}\,)$ is the particle 4-momentum, $f_a(x,p)$ represents the distribution function of the particle of type `$a$' and $g_a$ is the corresponding spin-isospin degeneracy factor, $\rmd\sigma_\mu=n_\mu\rmd^3\sigma$ is an element of the space-time freeze-out hypersurface with the normal  $n_\mu$.
The freeze-out hypersurface, as in the previous work~\cite{HYDHSD2015}, is determined with the help of the CORNELIUS algorithm~\cite{Huovinen}.

In the ideal-fluid case, the particle distribution function is given by usual the Fermi/Bose distribution
\begin{align}
f_{a}^{(0)}(x,p) &= \frac{1}{e^{\beta(p^\nu u_\nu(x)-\mu_a(x))} \pm 1}\,,
\end{align}
where $\beta=1/T$ is the inverse local temperature, $\mu_a$ is the chemical potential of the particle of type $a$ (Recall that the Coulomb interaction is neglected and all particles within in a given isospin multiplet have the same chemical potential). The plus and minus signs correspond to fermions and bosons, respectively.
For viscous fluids, one has to  take into account the modification of the distribution function because of non-equilibrium viscous effects
\begin{align}
f_a(x, p)=f_{a}^{(0)}(x,p)+\delta f_a(x,p)
\label{f-CF-full}
\end{align}
The common way is to approximate the viscous correction to the distribution function by the following expression~\cite{KHPB,AMY00,Teaney03}:
\begin{align}
\delta f_a(x,p)&=f_{a}^{(0)}(x,p)(1\mp f_a^{(0)}(x,p))\frac{p^\mu_a p^\nu_a\pi_{\mu\nu}}{2T^2(\varepsilon+P)}.
\label{visc-df}
\end{align}
The problem is that Monte-Carlo sampling needs a positive defined distribution and so the regions of negative $f_a(x, p)$ must be cut out~\cite{MEH_is3d}. Therefore one has to regulate the correction as
\begin{align}
\delta f^{\rm +}_a(x,p)&= \delta f_a(x,p)\Theta\big(f_{a}^{(0)}(x,p)+\delta f_a(x,p) \big),
\label{viscdistr}
\end{align}
where $\Theta(x)$ is the Heaviside step-function.

\begin{figure}
\includegraphics[width=8.8cm]{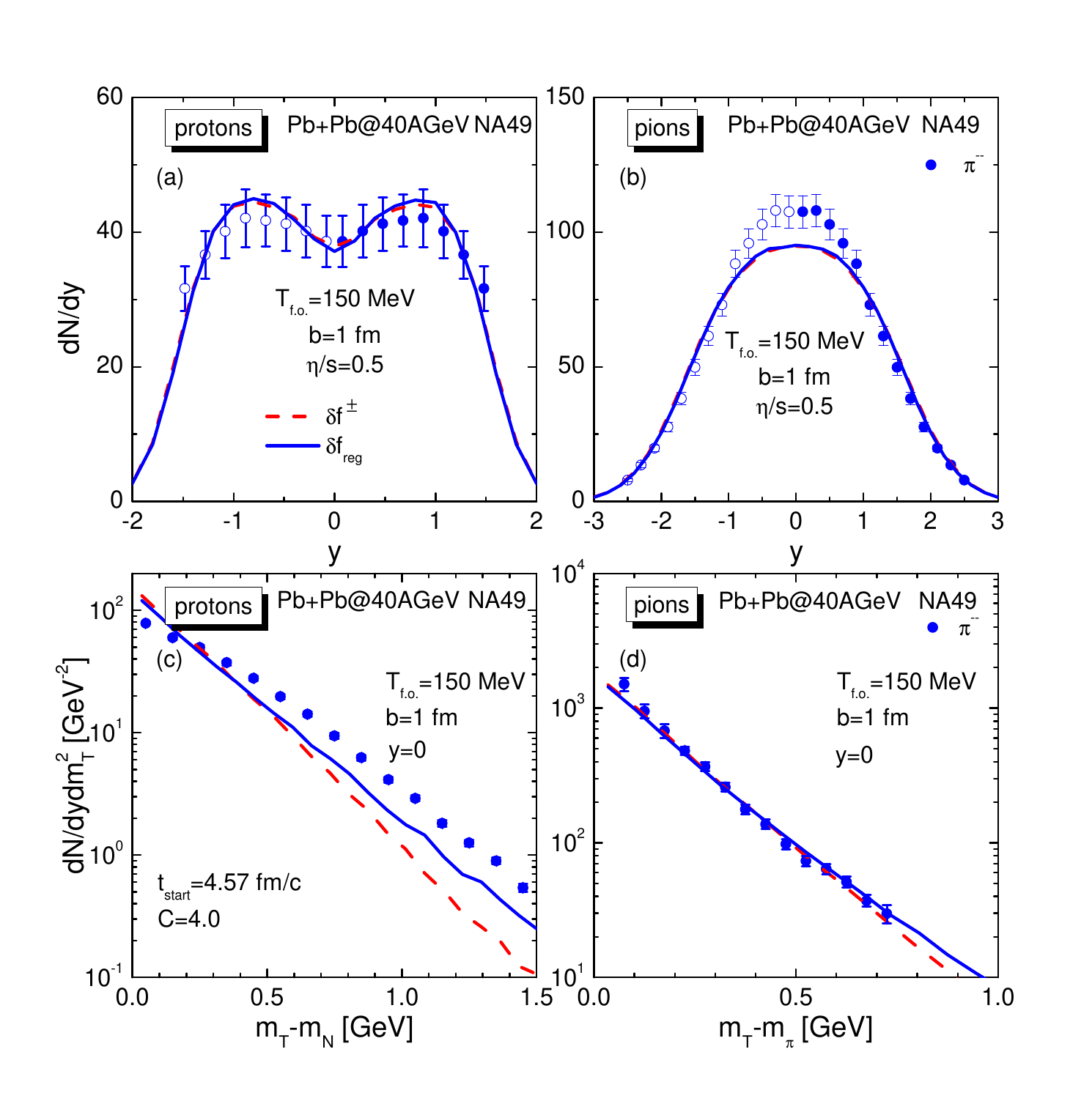}
\caption{Rapidity distributions and transverse momentum spectra for protons and pions produced in Pb+Pb collisions at 40\,$\agev$ in comparison with the calculations of the HydHSD model for various versions of the freeze-out algorithm and control of the viscous correction term (\ref{visc-df}). Solid lines show the results obtained with the cut prescription (\ref{viscdistr1}), dashed lines correspond to the regularization (\ref{freg}). The freeze-out temperature is fixed at $\Tfrz=150$\,MeV, $\eta/s=0.5$, the transition time is $t_{\rm start}=4.57\,{\rm fm}/c$ and the parameter $C=4$ in the condition (\ref{piconstrain}). Experimental points are taken from Refs.~\cite{SPSN1,SPSN2,SPSpiK}.
}
\label{fig:40AGeV-particlization}
\end{figure}

Another peculiarity of the particlization procedure for non-ideal hydrodynamics is that to apply the rejection procedure, one has to know the upper limit of the viscous correction factor. As seen, $\delta f_a(x,p)$ can become arbitrarily large at high momenta. The number of particles with too large values $\delta f_a$ depends in real calculations on the value of shear viscosity, the type of a constraint imposed by the inequality (\ref{neareqcondition}), the value of parameter $C$ in~(\ref{piconstrain}), and so on. This problem is common for different viscous hydrodynamical codes, see \cite{MEH_is3d}.  It is reasonable to assume that the viscous term has to be only a small correction, $|\delta f_a|<f_{a}^{(0)}(x,p)$. Let us mention that only in this case the temperature and chemical potential can be defined~\cite{deGroot}. There are two ways to fulfil this condition. The first one is to reject a particle momentum if
\begin{align}
\label{viscdistr1}
f_{a}^{(0)}(x,p)<|\delta f_a(x,p)|.
\end{align}
We will denote the such correction term as $\delta f^{\pm}_a(x,p)$.
Another way, proposed in Ref.~\cite{MEH_is3d} is to perform sampling with the regularized correction
\begin{align}
\delta f^{\rm reg}_a=\min\big[\delta f^{\rm +}_a, f_{a}^{(0)}(x,p)\big].
\label{freg}
\end{align}
The latter approach leads to an asymmetry: momenta, for which $\delta f<-f_{a}^{(0)}(x,p)$, must be forbidden at all, while the momenta, for which $\delta f>f_{a}^{(0)}(x,p)$, are just suppressed. In contrast, $\delta f^{\pm}_a(x,p)$ is a symmetric solution in the above mentioned sense. To estimate effects of different approaches, in Fig.~\ref{fig:40AGeV-particlization} we compare the results obtained using $\delta f^{\rm reg}$ and with the strict rejection. We chose a higher value of the regulator parameter $C$ entering the constraint (\ref{piconstrain}), $C=4$, to facilitate the possible effects of viscosity.
We see that the difference in the final transverse momentum spectra of protons and pions is comparable with the effect of increasing $\eta/s$ but is negligible for rapidity distributions.
Below we use $\delta f^{\rm reg}_a$ regularization. Therefore the ratio $\delta f^{\rm reg}_a/f_{a}^{(0)}(x,p)$ is not larger than 1.

We use exactly the same method of particle momentum generation~\cite{PSBBS,FastMC} as described in \cite{KHPB}. To use it, one has to convert $\pi^{\mu\nu}$ to the LRF. Due to the orthogonality relations~(\ref{orthogonality}) are explicitly fulfilled in our code, we have $\pi^{*0\nu}=0$ where the asterisk refers to the LRF.

After generating ``thermal'' contributions, 
resonance decays are taken into account in the zero-width approximation. To calculate the proton fraction among nucleons, we use isospin factor $1/2$ while for pions $1/3$.

\subsection{Equation of state}
\label{eos_sec}

The used EOS~\cite{SDM09} includes all known hadrons with masses up to 2\,GeV in the zero-width approximation.  The equation of state of hadron resonance gas at finite temperature and baryon density is calculated thermodynamically taking into account a density-dependent mean field that guarantees the nuclear matter saturation.

To account for mean-field effects, an effective potential $U = U(n)$ acting on a baryon is introduced. It depends only on the baryon density, $n$, and does not depend on the momenta of interacting baryons. Then the baryon’s single-particle energy can be obtained simply by adding $U(n)$ to the kinetic energy. In this case, the partition function of the hadronic system can be calculated analytically~\cite{PRD01}. As the result, the following expressions for
thermodynamic functions of the hadron EoS can be written:
\begin{align}
P &= \sum_a P_a(T,\mu^*,\mu_S) + P_f (n),
\\
\varepsilon &= \sum_a\varepsilon_a(T,\mu^*,\mu_S) + \varepsilon_f (n),
\end{align}
where the effective baryon chemical potential, $\mu^*$, is obtained by the shift $\mu^* = \mu_B - U(n)$.
The ``field'' contributions (marked by index 'f') to the
energy density and pressure are found as
$$
\varepsilon_f(n) = nU(n) - P_f (n) =\int_0^n \rmd n' U(n').
$$
In this approach, meson contributions are given by ideal gas expressions.

The mean-field potential is parameterized in a line with the Skyrme approach as $U(n) =\alpha n/n_0+ \beta(n/n_0)^\gamma$, where $n_0$ is the saturation density of nuclear matter
and $\alpha, \beta, \gamma ={\rm const}$. In the following, we fix $\gamma= 7/6$
and choose the remaining parameters from the requirements $P(T=0,n_0) = 0,\ \varepsilon(T=0,n_0)/n_0 = E_b+m_N$ where the binding energy $E_b=-16$\,MeV and $n_0=0.15\,{\rm fm}^{-3}$.
For more details on the EOS, see Ref.~\cite{SDM09}.

In the present study, we refrain from additional tunings of the EOS and the initial state.
We try to explain experimental data using only hadronic EoS to find observables, which cannot be described by a simple refitting of hydrodynamical parameters. Another reason is that changing EoS gives the additional very flexible degree of freedom and, in our opinion, should be used when the model parameter space will be well investigated.

\section{Influence of model parameters on momentum spectra}
\label{param_depend}

In this section, we consider how a variation of the hydrodynamic model parameters can manifest itself in rapidity ($y$) distributions and transverse momentum ($m_{\rm T}$) spectra at  $y=0$ of protons and pions. To be specific, we consider Pb+Pb collisions at 40\,$\agev$.

\subsection{Shear viscosity}

\begin{figure}
\includegraphics[width=8.8cm]{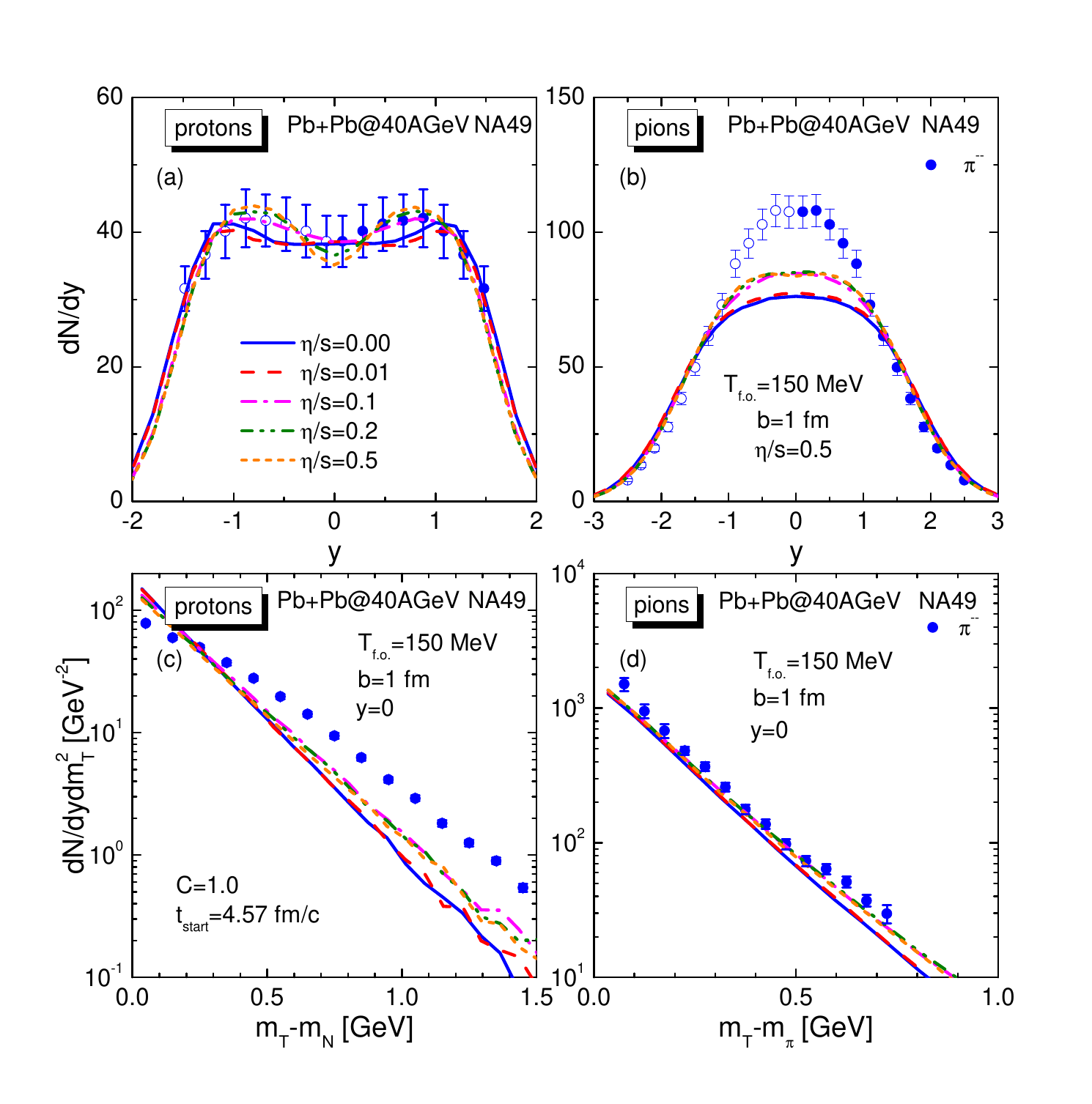}
\caption{Rapidity distributions and transverse momentum spectra for protons and pions produced in Pb+Pb collisions at 40\,$\agev$ in comparison with the calculations of the HydHSD model for various values of the $\eta/s$ ratio. The freeze-out temperature is fixed at $\Tfrz=150$\,MeV, the transition time is $t_{\rm start}=4.57\,{\rm fm}/c$, and $C=1$.  Experimental points are taken from Refs.~\cite{SPSN1,SPSN2,SPSpiK}.
}
\label{fig:40AGeV-eta}
\end{figure}

First of all,  let us compare proton and pion $y$- and $m_{\rm T}$-distributions evaluated for viscous and ideal hydrodynamics. The results are collected in Fig.~\ref{fig:40AGeV-eta}. One expects that calculations with a very small value of $\eta/s =0.01$ have to be very close to ideal-hydro calculations. Figs.~\ref{fig:40AGeV-eta}a and~\ref{fig:40AGeV-eta}b demonstrate that it is true. One should take into account the systematic uncertainty ($\sim\pm 1$ nucleons) in the particle number because of the Monte Carlo particlization procedure. The mentioned error source is further included in errors of fitted parameters, see Table~\ref{tab:fit-parameters}, and does not influence qualitative conclusions, which we make below. For the $m_{\rm T}$ spectra at mid-rapidity, the difference between calculations with $\eta/s=0$ and $\eta/s=0.01$ is negligibly small, as we see in panels (c) and (d) of  Fig.~\ref{fig:40AGeV-eta}.
An increase in the viscosity up to $\eta/s=0.1$ leads to markable changes in the rapidity distributions, see the dashed lines in Figs.~\ref{fig:40AGeV-eta}a and~\ref{fig:40AGeV-eta}b.
For protons, the hump heights increase slightly and the positions are shifted towards midrapidity. It occurs because the shear viscosity slows the fireball longitudinal expansion and the fluid velocity. Simultaneously, the proton rapidity distribution narrows because of the baryon number conservation. The changes increases for larger viscosities, $\eta/s=0.2$ and $\eta/s=0.5$ shown by dash-dot-dot  and short-dash lines.
The dip at midrapidity increases and the two-hump structure in the proton rapidity distribution becomes more pronounced.
For pions, the viscous corrections make the pion rapidity distributions higher than in the ideal case, see Fig.~\ref{fig:40AGeV-eta}b, as was anticipated in Ref.~\cite{HYDHSD2015}.

The transverse momentum spectra of pions and protons show the weak dependence on the $\eta/s$ value, see Figs.~\ref{fig:40AGeV-eta}c and~\ref{fig:40AGeV-eta}d, especially for pions. The inclusion of viscosity leads to a slight increase of slopes of the $m_{\rm T}$ spectra that brings the spectra closer to experimental data. It can be considered as an additional argument for the necessity of non-zero shear viscosity.

In Fig.~\ref{fig:40AGeV-eta} we also see that the influence of the viscosity on the particle momentum distributions saturates for large value of $\eta/s$, so the lines calculated for $\eta/s=0.2$ and $0.5$ are almost indistinguishable. This is because of the strict constraint on the $\pi^{\mu\nu}$ tensor (\ref{piconstrain}) with (\ref{q-def-S}), which we apply in our calculations with $C=1$.
Thus the height of the pion rapidity distribution measured in experiments at $\Elab=40\agev$ cannot be reproduced by a further increase of the viscosity parameter.

\subsection{Freeze-out temperature}\label{ssec:Tfo}

\begin{figure}[h]
\includegraphics[width=8.8cm]{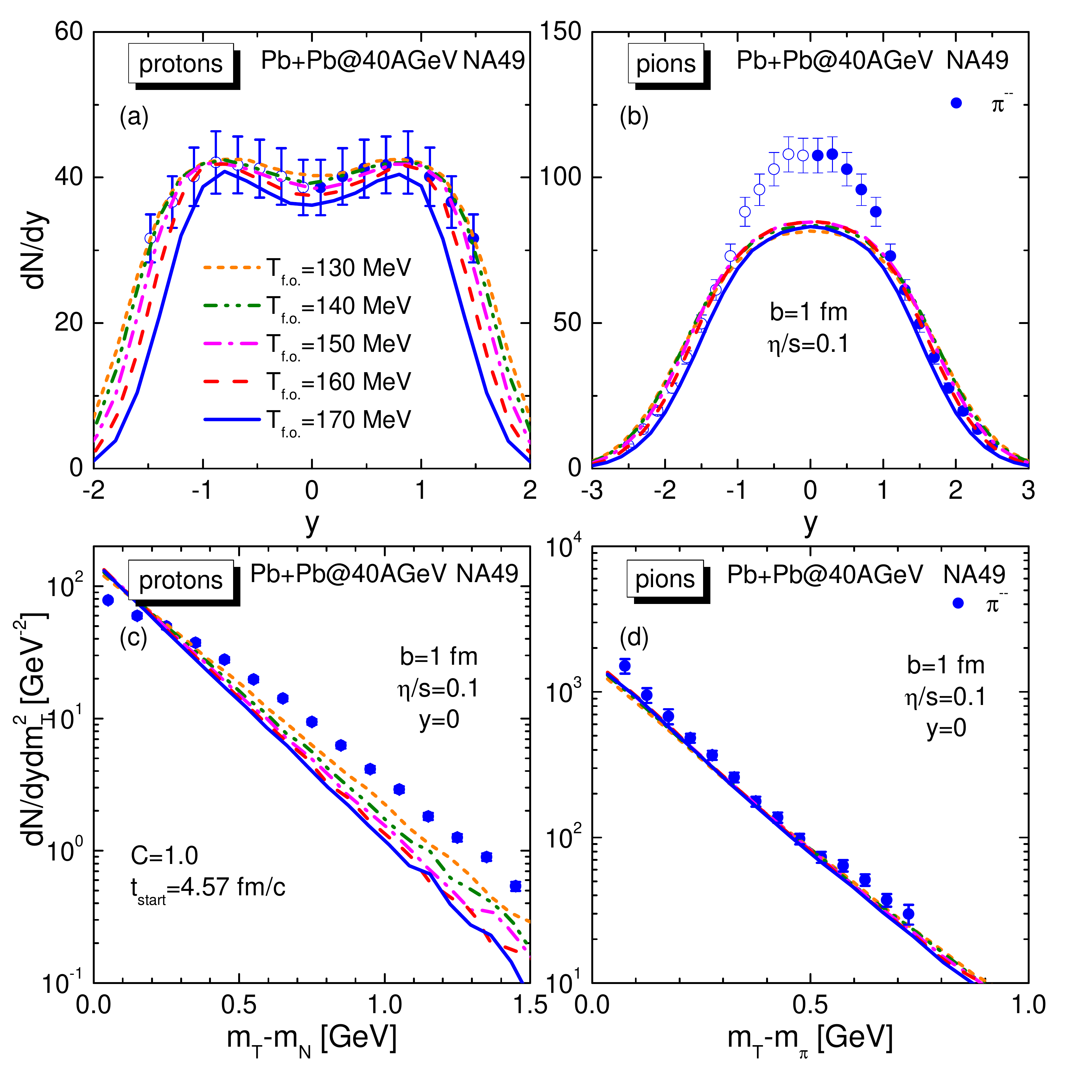}
\caption{
Rapidity distributions and transverse momentum spectra for protons and pions produced in Pb+Pb collisions at 40\,$\agev$ in comparison with the calculations of the HydHSD model for various values of the freeze-out temeperature $\Tfrz$. The viscosity is fixed at $\eta/s=0.1$, the transition time, the cut-off parameter, $C$, and experimental points are the same as in Fig.~\ref{fig:40AGeV-eta}.
} \label{fig:40AGeV-Tfo}
\end{figure}

The influence of the freeze-out temperature, $\Tfrz$, on rapidity distributions and transverse momentum spectra at mid-rapidity is illustrated in Fig.~\ref{fig:40AGeV-Tfo}. As is seen in the figure, the proton
rapidity distributions become slightly higher and wider if the freeze-out temperature $\Tfrz$ is lowered from 170\,MeV to 130\,MeV. The value of  $dN/dy$ at $y=0$ is moderately sensitive to $\Tfrz$ as well as to $\eta/s$, as we demonstrated above.
This value is sensitive to the transition time, $t_{\rm start}$, which determines the initial energy density and temperature distributions in the fluid, see Fig.~\ref{fig:t-start} and its discussion in the text.

The width of the pion rapidity distribution is larger for smaller values of $\Tfrz$, as is seen in Fig.~\ref{fig:40AGeV-Tfo}b, whereas the distribution height depends weakly on $\Tfrz$. Also, we observe saturation of the height of the pion $y$-distribution with decreasing $\Tfrz$ similar to the dependence on $\eta/s$. As can be seen in Figs.~\ref{fig:40AGeV-eta}b and \ref{fig:40AGeV-Tfo}b, the height of the pion rapidity distribution saturates at the level, which is significantly below the experimental data at the mid-rapidity.
Thus, if we vary only $\eta/s$ and $\Tfrz$ parameters within our standard calculation set up, we can reproduce
only the proton rapidity distribution but not the pion one for collisions at $\Elab=40\agev$.
Fig.~\ref{fig:40AGeV-Tfo} demonstrates that there is an internal tension in attempts to describe simultaneously the proton and pion rapidity distributions in our model. The origin of this problem is in the discussed-above insensitivity of the $y$-distributions to an increase of  $\eta/s$ above 0.1 value. Therefore, after the increase in the distribution by the variation of $\eta/s$ is exhausted, we have only one parameter $\Tfrz$ to tune both proton and pion distributions.  So for a freeze-out temperature, $130\,{\rm MeV}\lsim \Tfrz \lsim 160$\,MeV, which is needed to fit the proton rapidity distribution, we have only the correct width of the pion distribution.
The problem could be partially remedied if one included a finite width of resonances which will increase the population of low-momentum pions.

Transverse momentum spectra of protons and pions at mid-rapidity ($y=0$) are shown in Figs.~\ref{fig:40AGeV-Tfo}c and~\ref{fig:40AGeV-Tfo}d, respectively, for various values of freeze-out temperature. The striking feature is that the slope of the pion spectra is almost insensitive to the variation of $\Tfrz$ and the proton spectra demonstrate weak dependence on $\Tfrz$, whereby the slope steepness decreases with a decrease of $\Tfrz$. Hence, to approach experimental data for the proton $m_T$-spectrum we have to choose a quite low temperature in contrast to the statistical model \cite{Andronic}, see Section \ref{fitsection}.

\subsection{Constraints on the shear stress tensor}
\label{VHLLEMUSIC}

The above results lead to two questions. Why viscous effects in our 2-stage hybrid model for pion rapidity distribution are so small ($\sim$10\%) while the results of the authors~\cite{KHPB} within the vHLLE+UrQMD model demonstrate that the response is large (about 20\%, see Fig.~4 in Ref.~\cite{KHPB})? It cannot be explained by taking into account the electric charge conservation since this effect is included in both ideal and viscous versions of the model~\cite{KHPB}. The second question is why our model is insensitive to the $\eta/s$ value at $\Elab=40\agev$, if $\eta/s>0.2$ (see Fig.~\ref{fig:40AGeV-eta})?

\begin{figure}
\centering
\includegraphics[width=8.8cm]{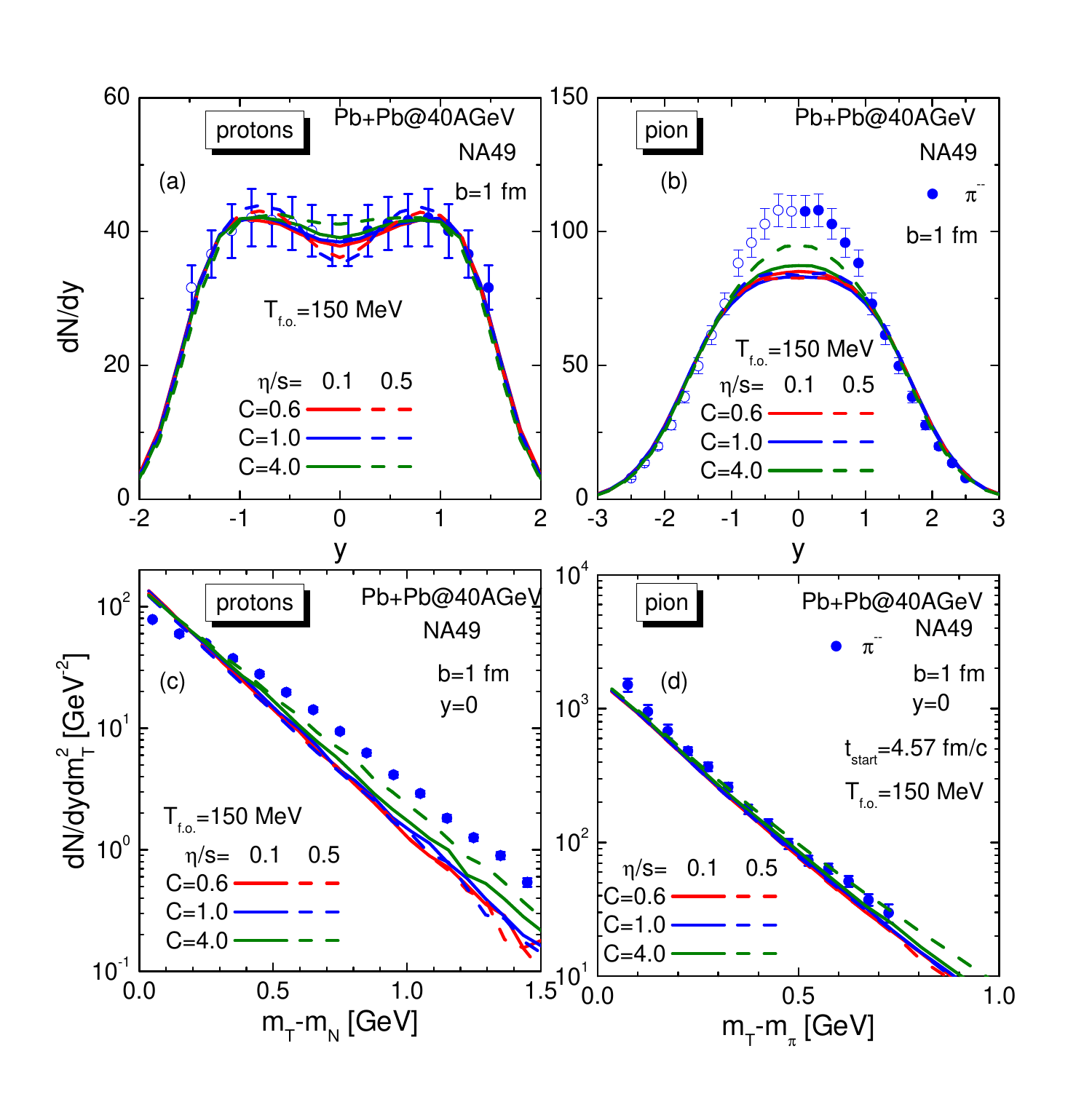}
\caption{
Proton and pion rapidity distributions and  $m_T$ spectra calculated for various values of the $C$ parameter in the shear-stress tensor constraint (\ref{piconstrain}) and (\ref{q-def-S}) for $\eta/s=0.1$ and $\eta/s=0.5$. The freeze-out temperature $\Tfrz$, the transition time $t_{\rm start}$, and experimental data are the same as in Fig.~\ref{fig:40AGeV-eta}.
} \label{fig:40AGeV-C}
\end{figure}

We anticipate this is related to the regularization of viscous effects by condition (\ref{piconstrain}) applied to assure that the viscous part of the energy-momentum tensor remains perturbative and the hydrodynamic equations, we solve, Eqs.~(\ref{hydrobase-T}), (\ref{hydrobase-J}) and (\ref{ISeqs}), keep their validity.

\begin{figure}
\centering
\includegraphics[width=8.8cm]{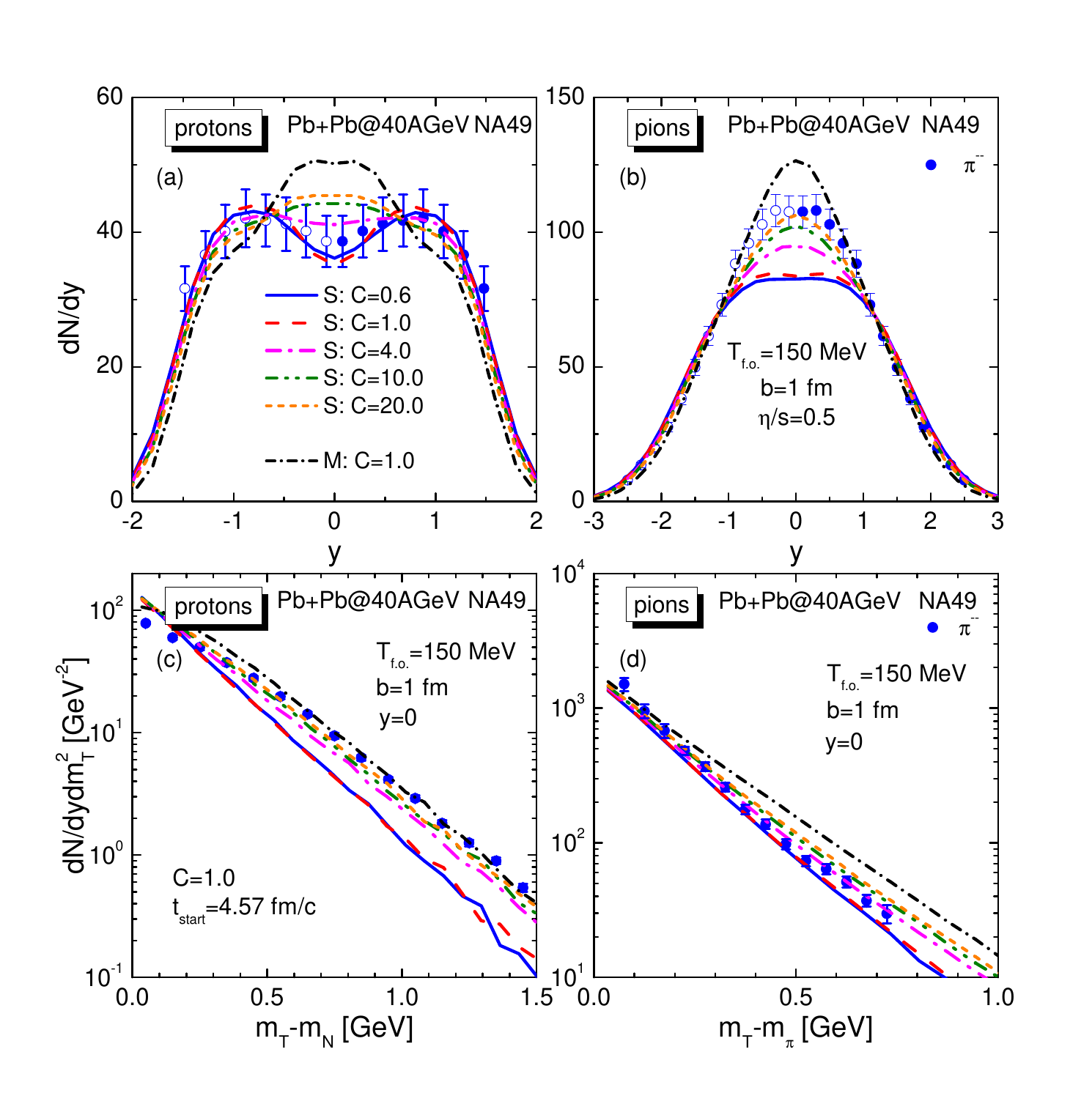}
\caption{
Proton and pion rapidity distributions and  $m_T$ spectra calculated for various values of the $C$ parameter in the shear-stress tensor constraint (\ref{piconstrain}) for $\eta/s=0.5$. For lines marked by 'S' the quantity $q$ is calculated as in Eq.~(\ref{q-def-S}), we call it the S-condition. Lines marked by 'M' are calculated with the M-constraint when the quantity $q$ is given by Eq.~(\ref{pimusic}).
The freeze-out temperature $\Tfrz$, the transition time $t_{\rm start}$ and experimental data are the same as in Fig.~\ref{fig:40AGeV-eta}.}
\label{fig:40AGeV-C-2}
\end{figure}

First, let us investigate how the viscous response changes if the regulator constant $C$ in Eq.~(\ref{piconstrain}) is increased. For larger $C$, the viscous effects are expected to be more pronounced. As Figs.~\ref{fig:40AGeV-C}, we compare the rapidity and $m_T$ spectra for protons and pions calculated for $\eta/s=0.1$ (solid lines) and $\eta/s=0.5$ (dashed lines) at increasing values of the regulator $C$ in  the condition (\ref{piconstrain}) for the S-constraint (\ref{q-def-S}). Comparing solid and dashed lines we see that with an increase of $C$ the difference among them increases. The height at mid-rapidity of proton and pion $y$-distributions is growing up significantly; humps in $dN_p/dy$ disappear at quite large values of $C$.  The reason for this strong increase will be discussed later in this section.
Also, the slope of the proton $m_T$ spectrum flattens. In Fig.~\ref{fig:40AGeV-C-2} we increase parameter $C$ further on (for $\eta/s=0.5$) and show the result for $C=10$ and 20. We see that the trends seen in Fig.~\ref{fig:40AGeV-C} continue: $dN/dy$ values at mid rapidity increase for both protons and pions. The shape of proton rapidity distribution changes from two humps to one hump at the midrapidity, and the height of the pion $y$-distribution reaches now the experimental date for $C=20$. The slope of the $m_T$ spectra flattens, and we fit the experimental data for $C=20$.

Second, different hydrodynamical models use different criterion to compare magnitudes of ideal, $T_{\rm id}^{\mu\nu}$ and viscous parts $\pi^{\mu\nu}$ of the energy momentum tensor. So, the vHLLE model~\cite{KHB2013} calculates the quantity $q$ as follows
\begin{align}
q=q_{\rm V}\equiv \frac{\max_{\mu,\nu} \left|\pi^{\mu\nu}\right|}{\max_{\mu,\nu}\left|T_{\rmid}^{\mu\nu}\right|}\,.
\quad \mbox{(V-cond.)}
\label{pivHLLE}
\end{align}
We will denote the condition (\ref{piconstrain}) with the quantity $q=q_V$ as the V-condition.
In the MUSIC model~\cite{MUSIC} the quantity $q$ is defined as
\begin{align}
\label{pimusic}
q=q_{\rm M}\equiv \sqrt{\frac{\pi^{\mu\nu}\pi_{\mu\nu}}{T_{\rm id}^{\mu\nu}T_{\rm id,\mu\nu}}}
\,. \quad \mbox{(M-cond.)}
\end{align}
We will call it the M-condition\footnote{As one can see from the function {\it QuestRevert} of MUSIC code or \cite{Denicol18}, the developers use an energy-dependent cut-off parameter $C=C(\varepsilon)$ in Eq.~(\ref{piconstrain}). We take just a constant value for simplicity. Our results do not changes if one takes $C(\varepsilon)\propto\tanh\frac{\varepsilon}{\varepsilon_0}$ with small $\varepsilon_0$.}.
The conditions (\ref{pivHLLE}) and (\ref{pimusic}) are more `integral' and, therefore, weaker than the strict (S-) constraint using $q$ defined in Eq.~(\ref{q-def-S}). They do not guarantee that each element of the viscous stress tensor $\pi^{\mu\nu}$ does not exceed the corresponding element of the ideal tensor.

Both V- and M- conditions can be easily realized in our code. In Fig.~\ref{fig:40AGeV-C-2} we show the results of calculations performed with the M-condition and the control parameter $C=1$. In this case, all effects associated with viscosity are proliferated even in comparison with the calculation for $C=20$ when the S-condition is applied. There clear bumps at mid-rapidity in proton and pion distributions going above the experimental points. The pion $m_T$ spectra also overestimate the experiment.

\begin{figure}
\centering
\includegraphics[width=8.8cm]{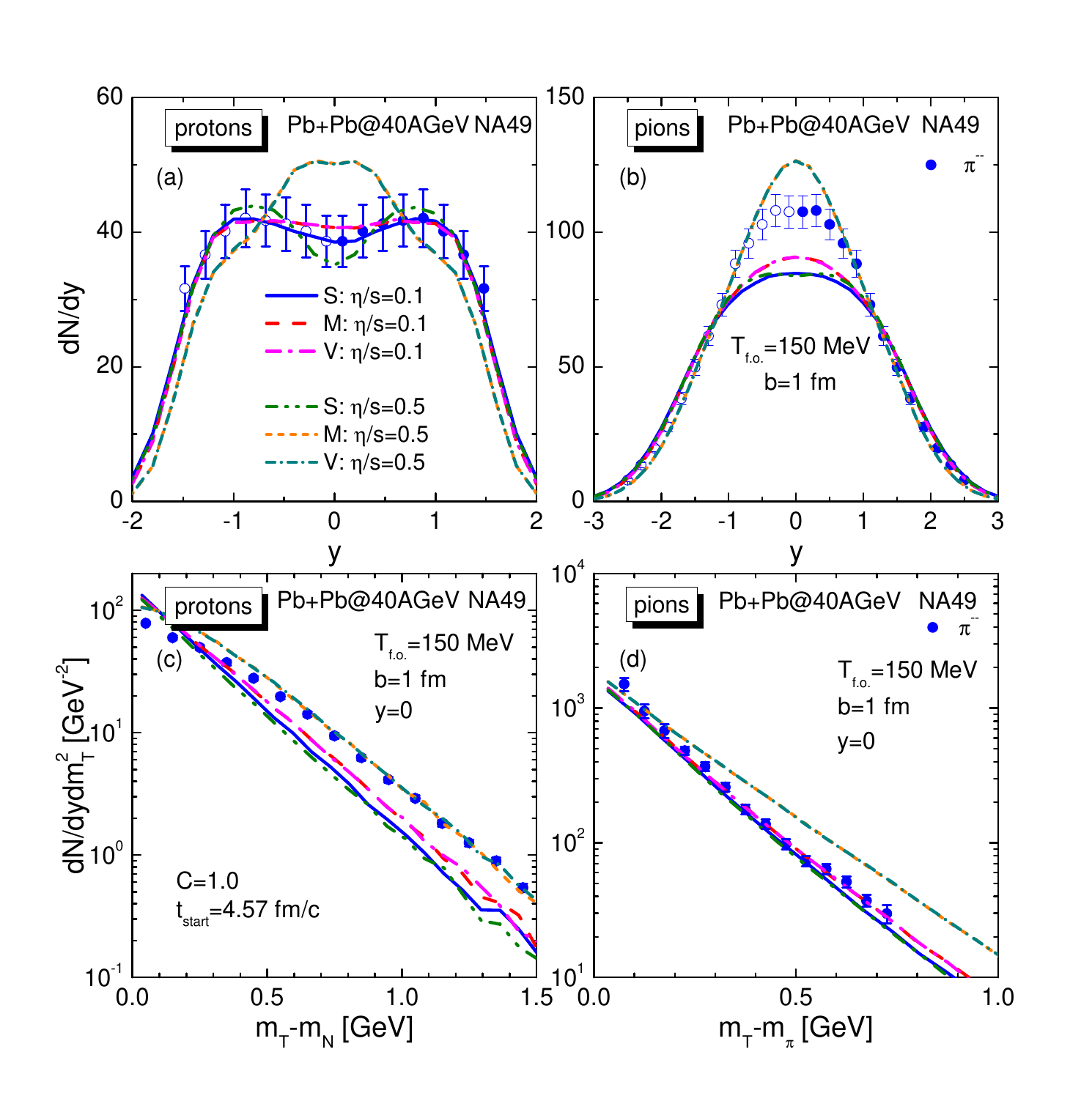}
\caption{Proton and pion rapidity distributions and  $m_T$ spectra for Pb+Pb collisions at $\Elab=40\,\agev$ calculated for different choices of $\pi^{\mu\nu}$ constraints (\ref{piconstrain}) with the parameter $q$ calculated according Eq.~(\ref{q-def-S}), lines marked by S, Eq.~(\ref{pivHLLE}), lines marked by V, and (\ref{pimusic}), lines marked by M. Calculations are carried out for the freeze-out temperatures $\Tfrz=150$\,MeV, transition time $t_{\rm start}=4.57\,{\rm fm}/c$ and two values of the shear viscosity $\eta/s=0.1$ and $0.5$, $C=1$. The experimental data are the same as in Figs.~\ref{fig:40AGeV-eta}-\ref{fig:40AGeV-C-2}.}
\label{fig:40AGeV-pi}
\end{figure}

A systematic comparison on S-, M-, and V-condition on the viscous stress tensor $\pi^{\mu\nu}$ is presented in Fig.~\ref{fig:40AGeV-pi}
for $\eta/s=0.1$ and 0.5 and the same value of parameter $C=1$. The difference of various conditions is only in how the quantity $q$ is calculated: according to Eq.~(\ref{q-def-S}) for the S-condition, Eq.~(\ref{pivHLLE}) for the V-condition, or Eq.~(\ref{pimusic}) for the M-condition. First of all, we conclude that the M- and V-conditions produce the same results for the rapidity and $m_T$ distribution for both protons and pions, as the corresponding lines are indistinguishable for both values of $\eta/s$.
This property is found to be valid for all collision energies considered in this work. So, everywhere below, speaking about the M-constraint we mean also the V-condition except the cases when noted separately.
Further, one can see that applying a weaker constraint (M and V) leads to almost no changes for $\eta/s=0.1$ but to a dramatic discrepancy in the rapidity distributions for larger viscosities. For $\eta/s=0.5$ the height of the proton humps and the maximum of the pion distribution increase sizably compared to the calculations with the stricter constraint (\ref{q-def-S}). Also, Fig.~\ref{fig:40AGeV-pi} shows that the sensitivity of the rapidity spectra to the $\eta/s$ value is much larger for the M-condition than for the S-condition.
We see in Fig.~\ref{fig:40AGeV-pi}b that with the weaker constraints one can reproduce the mid-rapidity height in the pion rapidity distribution by changing the $\eta/s$ parameter. This coincides with observations made in Ref.~\cite{KHPB}.
Figure~\ref{fig:40AGeV-pi} shows also that the results of calculations with the M-condition, Eq.~(\ref{pimusic}), coincide for $C=1$ and $C\geq1$. The same independence of the results for a variation of $C$ in the wide range  $C=1\mbox{--}30$ was found in Ref.~\cite{Denicol18}.
Thus, comparing these observations with the $C$ dependence of the results for S-constraint shown in Fig.~\ref{fig:40AGeV-C-2}, we may conclude that the M(V)-conditions with $C=1$ corresponds to the S-condition with $C\gg 1$, see also the M-results in Fig.~\ref{fig:40AGeV-C-2}. Many authors, see Refs.~\cite{MNR2010,KHB2013,DuHeinz19}, supposed that by taking $C\leq 1$ in their codes using the weaker M- and V-constraints they keep the viscous corrections small. But as we see, the smallness of shear viscous effects is not guaranteed unless the results are close to those obtained with the S-condition.

\begin{figure}
\centering
\includegraphics[width=7cm]{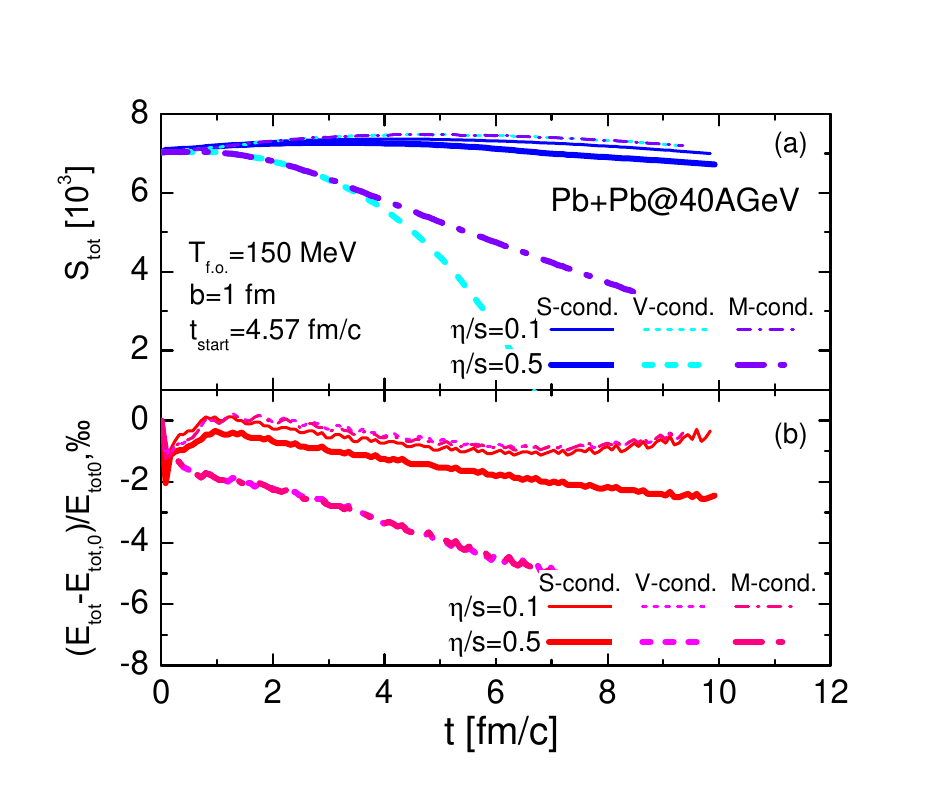}
\caption{Evolutions of the total entropy (panel a)  and total energy (panel b) for different choices of constraints on the viscous stress tensor $\pi^{\mu\nu}$ (S-, M-, and  V-constraints) for Pb+Pb collisions at $\Elab =40\,\agev$ with $\Tfrz=150$\,MeV for $\eta/s=0.1$ and $0.5$, and $C=1$.}
\label{fig:EStot-pi}
\end{figure}

Figure~\ref{fig:EStot-pi}a demonstrates that different constraints on the $\pi^{\mu\nu}$ tensor affect also the evolution of such a global quantity as the total entropy of the system. For $\eta/s=0.1$ energy and entropy stay constant with good precision in types of conditions. For $\eta/s=0.5$ the situation changes drastically: the total entropy of the system decreases strongly for the M- and, especially, for the V-condition. Such a sharp entropy change for weak constraints occurs because of a large increase of the number of cells where the viscous correction to the entropy flow exceeds the ideal part. Generally, we can calculate $S_{\rm tot}$ choosing a higher minimal temperature of cells included in the evaluation. Then, the variation of $S_{\rm tot}$ would be smaller. The necessity of such a fine tuning and the difference in the entropy evolution, $S_{\rm tot}(t)$, between the M- and V-conditions for cut temperatures 50-70\,MeV suggests that it is safer to use the M-condition. The total energy does not demonstrate large deviations since it is conserved with the good accuracy for any constraint and the difference lies within errors of the numerical simulation, see Fig.~\ref{fig:EStot-pi}b.

\begin{figure}\centering
\includegraphics[width=8.8cm]{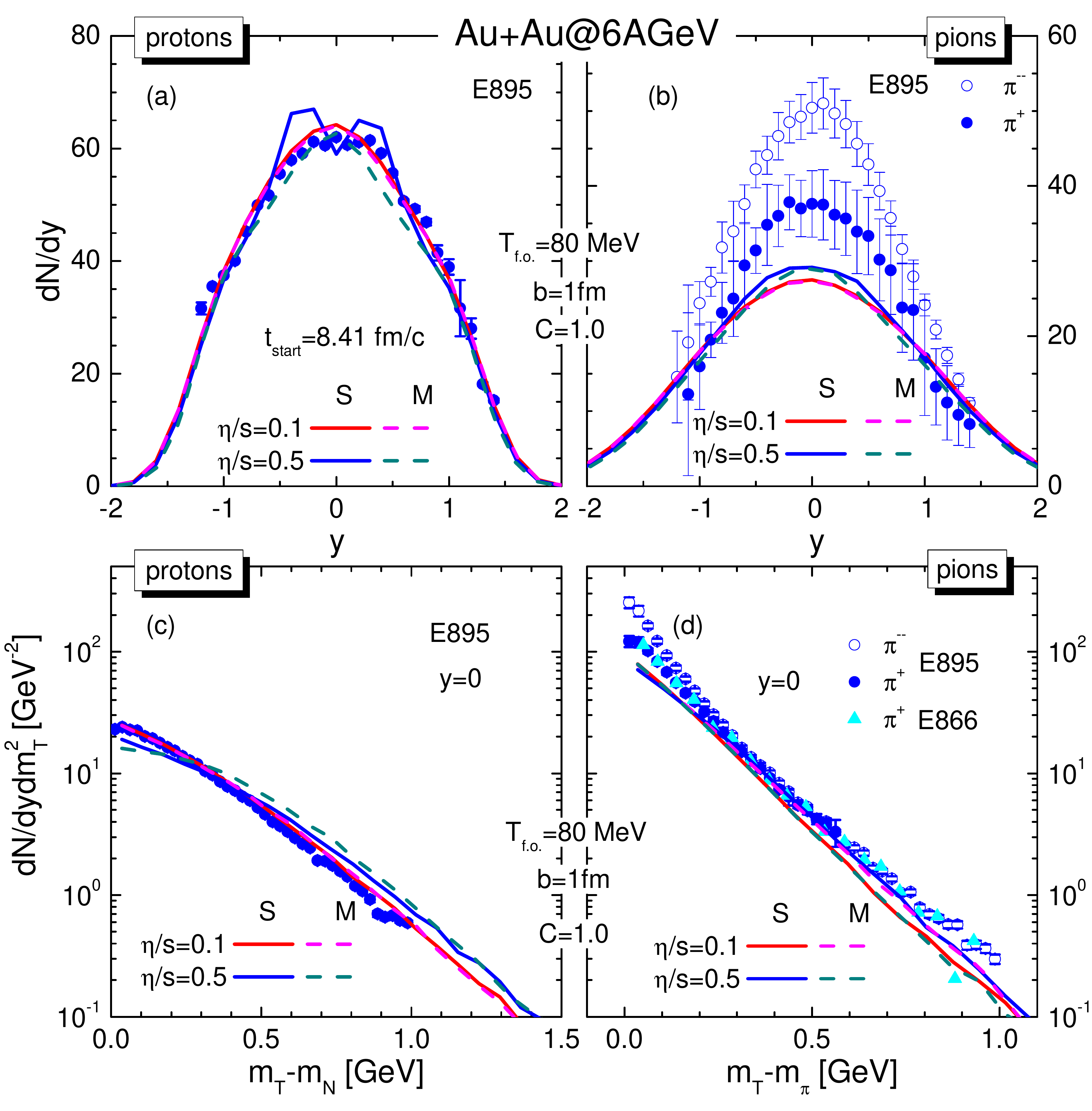}
\caption{ Rapidity distributions and $m_T$ spectra of protons and pions produced in Au+Au collisions at $\Elab=6\,\agev$ with $\Tfrz=80$\,MeV, transition time  $t_{\rm start}=8.41\,{\rm fm}/c$ and different conditions on the $\pi^{\mu\nu}$ tensor: the S-condition (\ref{q-def-S}) and the M-condition (\ref{pimusic}) shown be solid and dashed lines, respectively.
Results for two values of $\eta/s=0.1$, and $0.5$ are shown, $C=1$.
Experimental points are from~\cite{E895-prot,E895-pion}.}
\label{fig:6AGev-pi}
\end{figure}

To better understand how the form of the constraint on the shear stress tensor affects observables, let us consider also collisions at AGS energies. The results for Au+Au collisions at $\Elab=6\,\agev$ are shown in Fig.~\ref{fig:6AGev-pi}, where we put $t_{\rm start}=8.41$~fm/$c$, $\Tfrz=80$\,MeV and also consider two viscosities with $\eta/s=0.1$, and $0.5$. We use the standard value $C=1$. Also, we take into account nucleon coalescence as described in Ref.~\cite{HYDHSD2015}. As one can see, the S- and M-conditions give identical results for $\eta/s=0.1$.
The difference in proton quantities is enhanced for larger values of $\eta/s$: for the S-condition the two-hump structure $\rmd N_p/\rmd y$ develops, while for the M-condition two humps do not appear at large $\eta/s$. The maximum of the pion rapidity distribution increases with an increase of $\eta/s$. In general, the difference between the S- and M-conditions is larger for higher collision energies, compare Figs.~\ref{fig:40AGeV-pi} and \ref{fig:6AGev-pi} however the different conditions give different qualitative behavior of proton rapidity distributions with increasing $\eta/s$.

The obtained results confirm our earlier conclusion in Ref.~\cite{HYDHSD2015} that a two-hump structure in proton and pion distributions may have a kinematic (dynamic) origin and is not necessarily related to a phase transition.

\begin{figure}
\centering
\includegraphics[width=8.8cm]{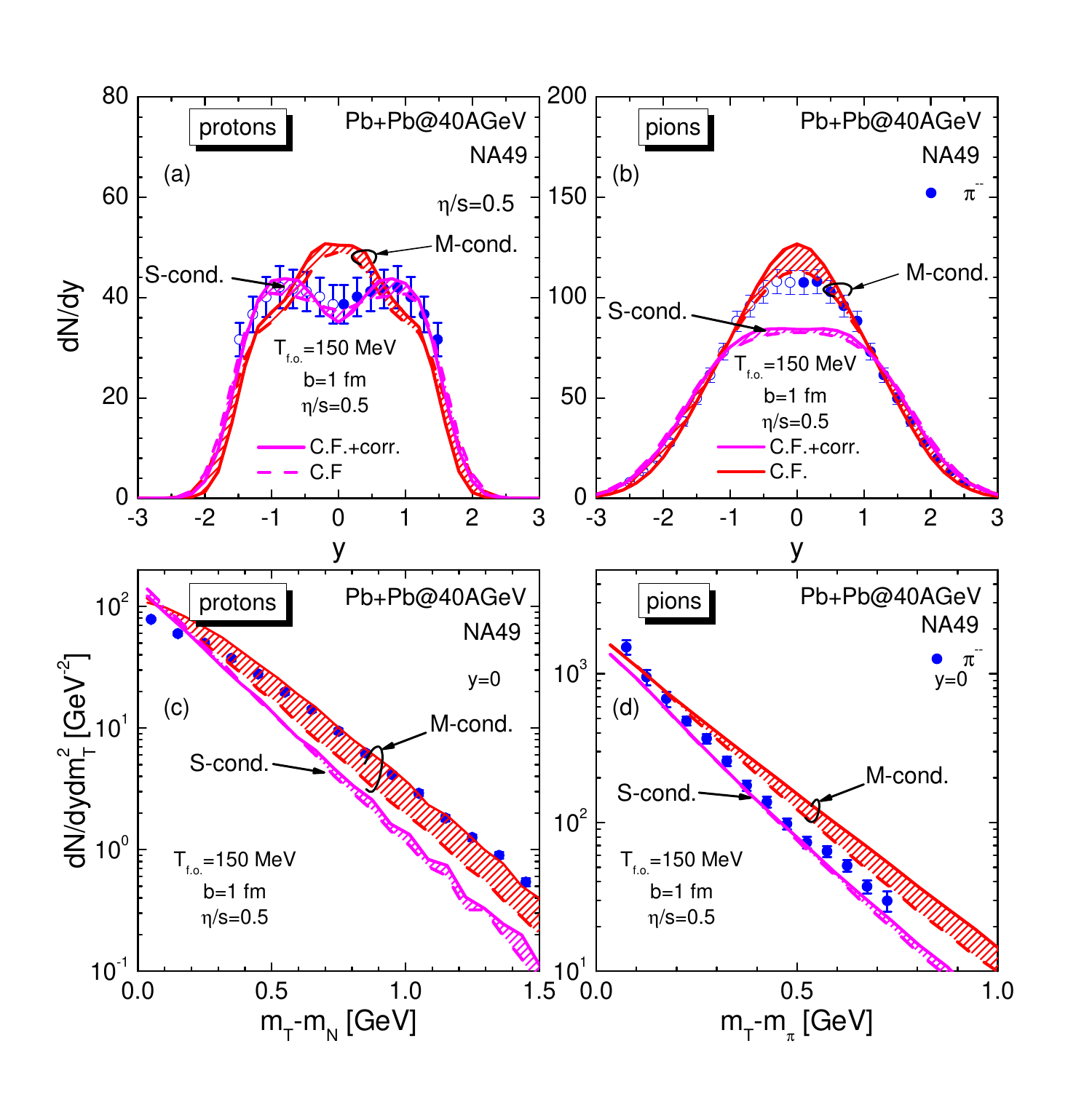}
\caption{Proton and pion rapidity distributions calculated using the Cooper-Frye formula
(\ref{CooperFrye}), (\ref{f-CF-full}) with and without viscous corrections (\ref{freg}) shown by solid and dashed lines, respectively. Calculations are performed for Pb+Pb collisions at $E_{\rm lab}=40\,\agev$ with $\eta/s=0.5$ for the S-condition and the M-condition, in both cases $C=1$. The freeze-out temperature $\Tfrz$, the transition time $t_{\rm start}$, and experimental data are the same as in Fig.~\ref{fig:40AGeV-eta}. }
\label{fig:CF-corr}
\end{figure}

In Fig.~\ref{fig:CF-corr} we illustrate the role of the viscous correction term in the Cooper-Frye formula (\ref{CooperFrye}) with (\ref{f-CF-full}) and (\ref{visc-df}) for Pb+Pb collisions at $E_{\rm lab}=40\,\agev$. Calculations are done for $\eta/s=0.5$. The regularization prescription (\ref{freg}) is applied. In panels (a) and (b), we present the proton and pion rapidity distributions calculated with and without the last term in square brackets in (\ref{viscdistr}), solid and dashed lines respectively. For pions the effect due to the correction term is sizably stronger for the M-condition.
For protons, the viscous correction term contributed with the same magnitude for both S- and M-conditions. Also for the S-condition, a noticeable correction to the humps is observed, that is natural since these rapidity regions are most sensitive to the viscosity.
For $m_T$ spectra shown in panels (c) and (d) the correction term is responsible for flattening of spectra observed previously in Figs.~\ref{fig:40AGeV-C-2} and \ref{fig:40AGeV-pi}. The effect is most pronounced for the M-condition.

\begin{figure}
\centering
\parbox{4cm}{\includegraphics[width=4cm]{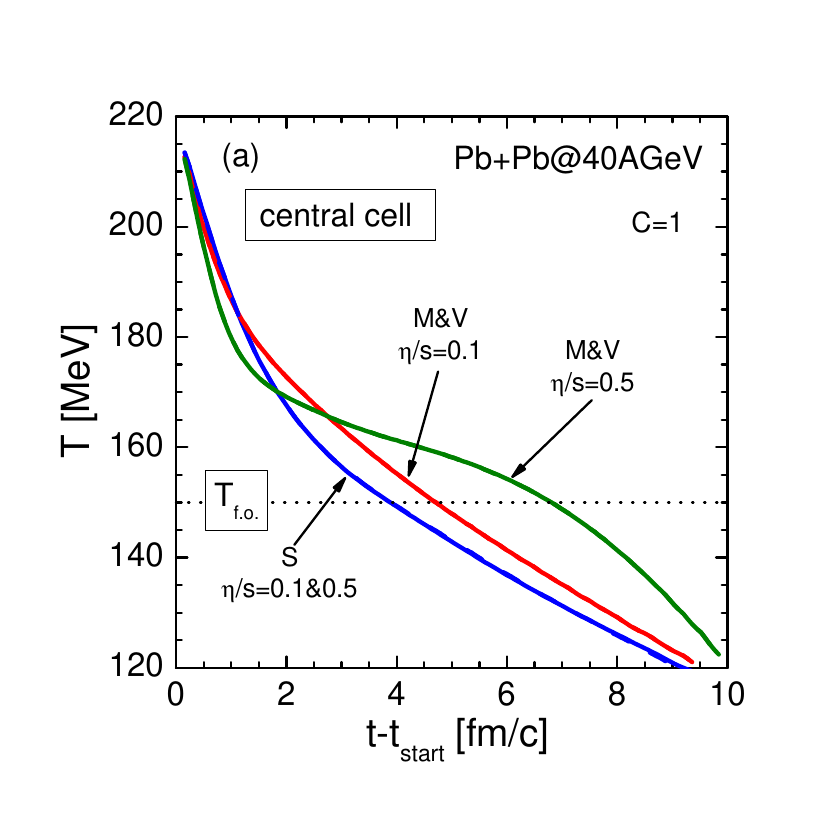}}
\parbox{4,05cm}{\includegraphics[width=4.05cm]{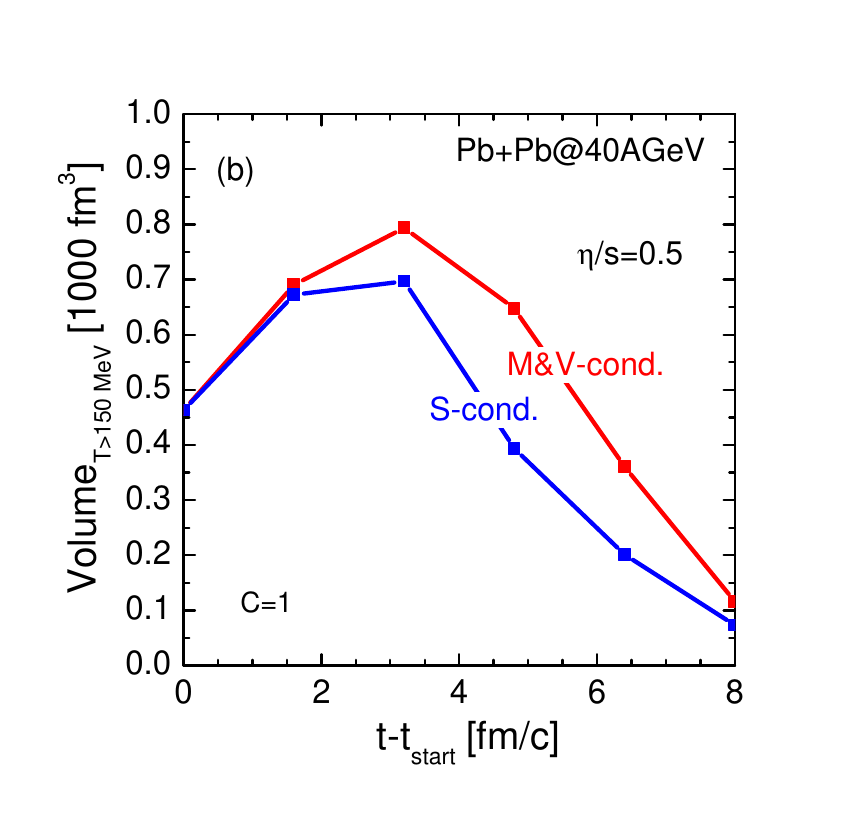}}
\caption{(a)~Evolution of the temperature in the central cell for various contraints and two values $\eta/s=0.1$ and $\eta/s=0.5$.
(b) The evolution of the volume of cells with temperatures $T>T_{\rm f.o.}=150$\,MeV for $\eta/s=0.5$.
Calculations are performed for Pb+Pb collision at $E_{\rm lab}=40\,\agev$ and  various conditions constraining the shear stress tensor: the S-, M-, and V-conditions. Time is counted from the transition time $t_{\rm start}=4.57\,{\rm fm}/c$.
}
\label{fig:centcell}
\end{figure}

Why the viscous effects promoted by the weak constraint with the V- and M-conditions lead to an increase in the pion number multiplicity? To address this question we present in Fig.~\ref{fig:centcell}a the evolution of the central-cell temperature for calculations done with various conditions. The viscous effects prolong the evolution and increase the temperature.  Also the number of fluid cells with temperatures $T>T_{\rm f.o.}=150$\,MeV increases for runs with weaker conditions (M and V), as illustrated in Fig.~\ref{fig:centcell}b. An increase in the specific viscosity results in a further increase in the volume. Thus, a combination of higher temperatures and larger freeze-out volume leads to a strong increase in the number of produced pions (not restricted by any conservation law) if the viscous effects are constrained by the V- and M-conditions.

\begin{figure*}
\centering
\includegraphics[width=5.25cm]{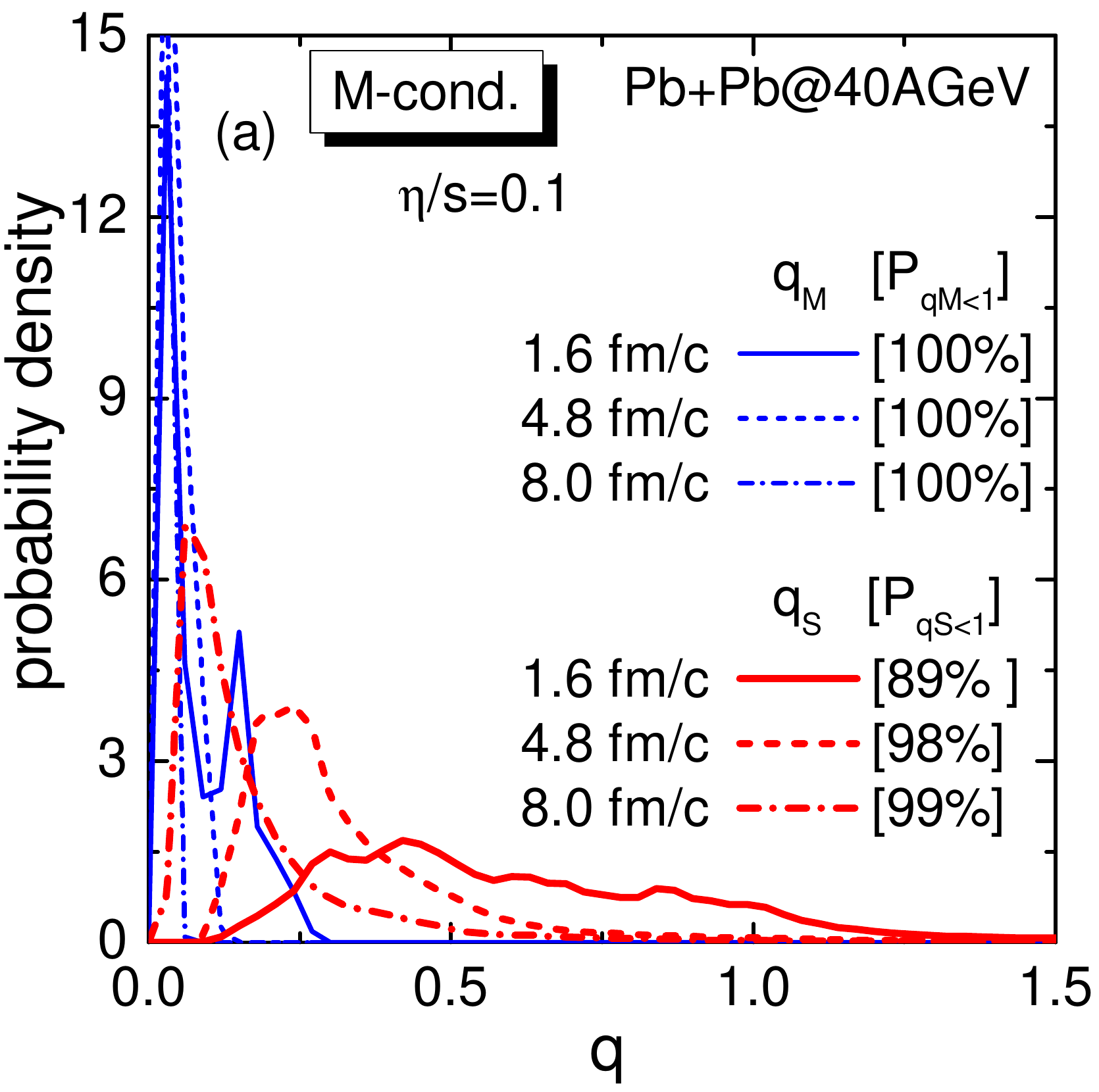}
\includegraphics[width=5cm]{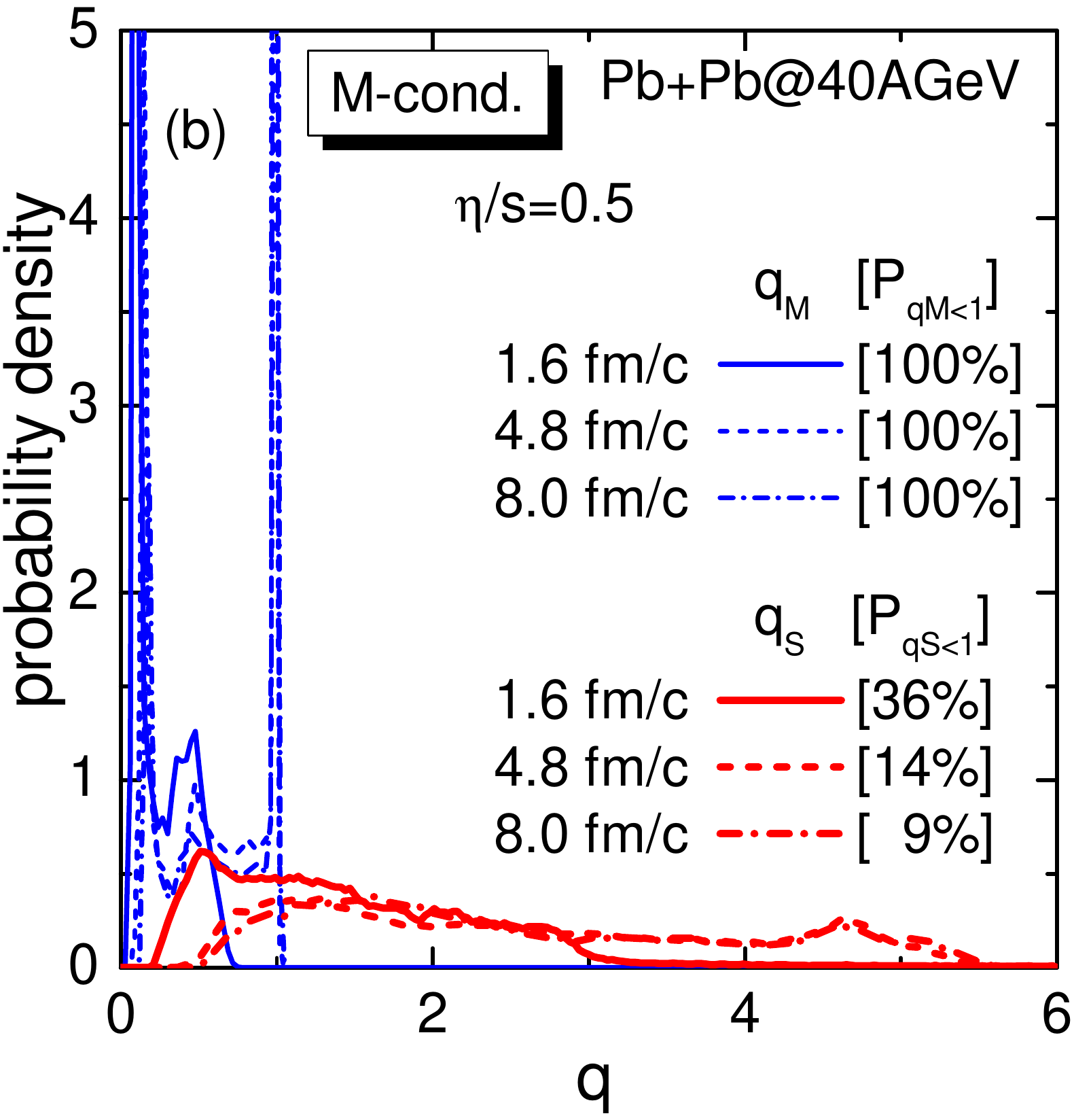}
\includegraphics[width=5cm]{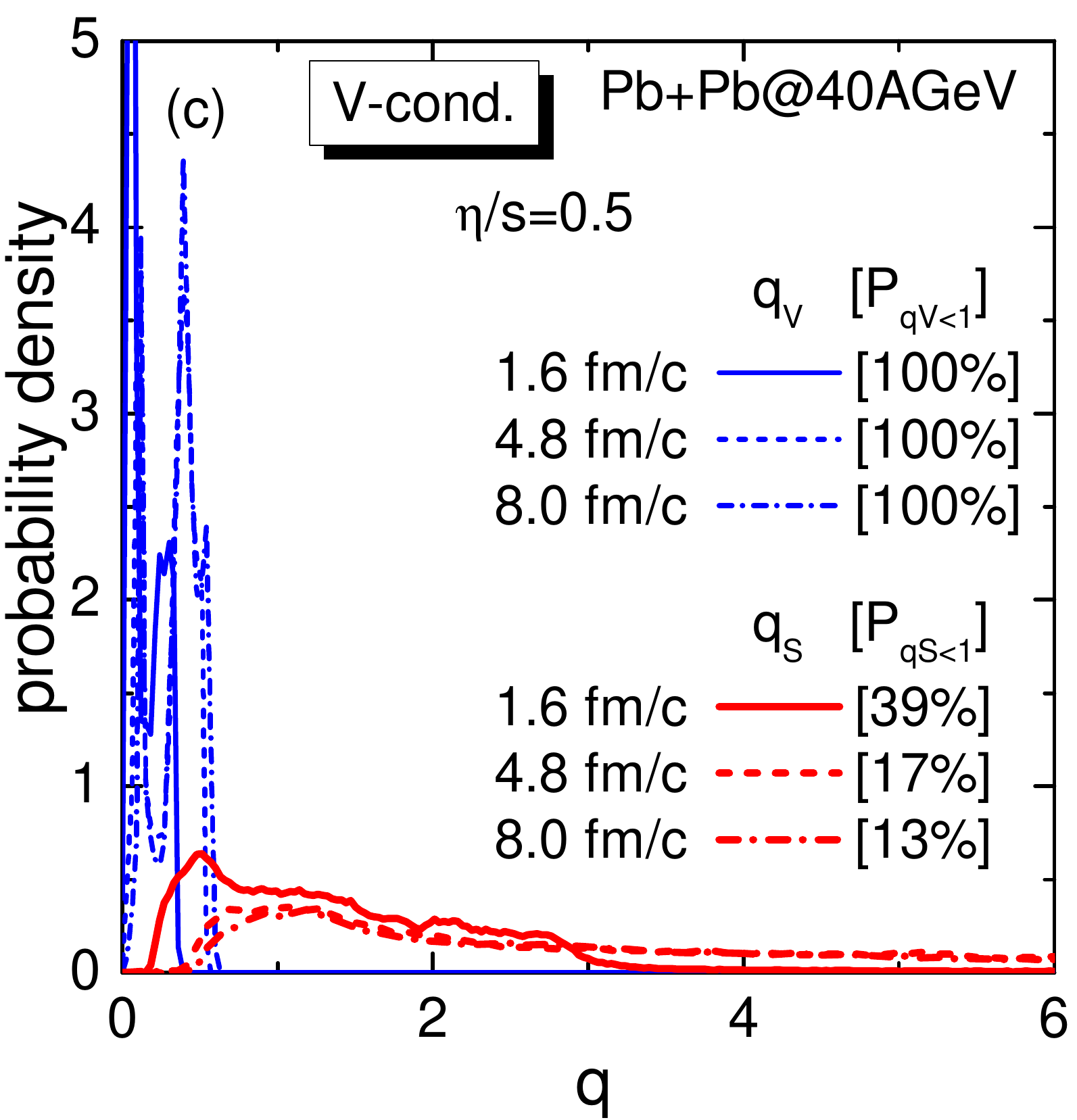}
\caption{Panel (a): probability density to find a fluid cell in the system with particular values of $q_{\rm M}$ (thin lines) and $q_{\rm S}$ (thick lines) parameters defined by Eqs.~(\ref{pimusic}) and~(\ref{q-def-S}) in hydrodynamic runs for the Pb+Pb collisions at $E_{\rm lab}=40\,\agev$ performed with the M-condition and $\eta/s=0.1$. Panel (b): the same as in panel (a) but for $\eta/s=0.5$.
Panel (c): the same as in panels (a) and (b) but for the run performed with the V-condition (\ref{pivHLLE}) and $\eta/s=0.5$.
On each panel, lines are shown for three values of $t-t_{\rm start}=1.6$, 4.8, 8.0\,fm/$c$.
Numbers in the square brackets show the probability to find a cell with $q_{\rm M,V,S}<1$.
}
\label{fig:q-distrib}
\end{figure*}

It is interesting to try to quantify to what extend the viscous effects remain perturbative in the course of hydrodynamic evolution. With this aim we run the code for Pb+Pb collision at $E_{\rm lab}=40\,\agev$ applying M- and V-conditions for $\eta/s=0.1$ and $\eta/s=0.5$ and calculated the distribution of the values of $q_{\rm M}$ and $q_{\rm V}$ defined in Eqs.~(\ref{pimusic}) and (\ref{pivHLLE}) respectively, among all fluid cells with temperatures $T>100$\,MeV. Then, in the same runs we calculated $q_{\rm S}$ defined in Eq.~(\ref{q-def-S}). In all runs we kept $C=1$. Let us now compare the distributions of various $q$ values. In  Fig.~\ref{fig:q-distrib}a we show the $q_{\rm M}$- and $q_{\rm S}$-distributions (thin and thick lines, respectively) for $\eta/s=0.1$ at three time moments. The distributions are normalized within the interval $0\le q\le 2$. We see that the condition (\ref{piconstrain}) is indeed respected during the code evolution: for all moments of time $q_{\rm M}$ remains safely below 1. On the other hand, the quantity $q_{\rm S}$ has a much wider distribution, however, it does not go far above 1. For early time, 1.6\,fm/$c$, 89\% percents of cells have $q_{\rm S}<1$ for later times this percentage increases. Thus we may conclude, for such a low value of $\eta/s$, the code runs in the regime when viscosity is just a perturbative correction. This conclusion holds also if we impose the V-condition instead of the M-condition.

The picture changes if we take larger value of $\eta/s=0.5$. Distributions of $q_{\rm M}$ and $q_{\rm S}$ values for the code running with the M-condition are shown in Fig.~\ref{fig:q-distrib}b and distributions of $q_{\rm V}$ and $q_{\rm S}$ for the run the V-condition in Fig.~\ref{fig:q-distrib}c. The distributions shown in this panels are normalized within interval $0<q<10$.
In both cases, the code keeps the values $q_{\rm M}$ and $q_{\rm V}$ smaller than 1.
The true characteristic for perturbativity of the viscous effect is, however, the quantity $q_{\rm S}$ calculated according to Eq.~(\ref{q-def-S}). If $q_{\rm S}>1$ then at least one of the elements in the $\pi^{\mu\nu}$ tensor is larger than the corresponding element in $T_{\rm id}^{\mu\nu}$, i.e. the viscous effect is non-perturbative and the applicability of the hydrodynamic equations (\ref{hydrobase}) and (\ref{ISeqs}) is questionable. The distributions of values $q_{\rm S}$ are shown in Figs.~\ref{fig:q-distrib}b and ~\ref{fig:q-distrib}c by thick lines.
We see that although the code is keeping $q_{\rm M, V}<1$ at each evolution step, the vast majority of fluid cells have $q_{\rm S}>1$. Although, we show only interval $0<q<6$ the distributions extend up to $q\simeq 10$. So, already at initial steps only in 36\% for the M-condition and 39\% for the V-condition of all cells, the viscous effects are truly perturbative. With time passed these numbers drop further down to even smaller values: 9\% and 13\%, respectively, at $t-t_{\rm start}=8\,{\rm fm}/c$.

Thus, applying the weak M- and V-conditions with $C=1$ at $\eta/s = 0.5$ we let the hydrodynamic code run, in reality, in the non-perturbative regime. (One should note that a weakening of constraint on the viscous term similar to the M(V)-condition can be also obtained for the S-condition if one lets $C$ be quite large.) This indicates that one cannot use the Israel-Stewart equations (\ref{ISeqs}), and one needs to include higher-order gradient terms on their r.h.s. This extension we will consider separately elsewhere. Also in such a regime, it might be necessary to take into account additional terms in Eq.~(\ref{viscdistr}). For the S-condition, the situation is different. We observe that rapidity distribution becomes almost insensitive to an increase of $\eta/s$ for $\eta/s>0.1$.  Therefore, we tend to rely more on the results, which are similar for both constraints, whereas the results, which are distinct, need separate investigations. As we will see in the next section, only the fit at $\Elab=10.7\ \agev$ gives large discrepancies.

\section{Beam-energy dependence of parameters}
\label{fitsection}

\begin{table*}[!ht]
\caption{The fitted parameters for Eq. (\ref{piconstrain}) (S-condition) and for Eq. (\ref{pivHLLE}) if proton [(p)V\&M-condition] or pion [($\pi$)S-condition] rapidity distribution are tuned. The fit accuracy is $\Delta T=\pm 5$\,MeV, $\Delta \eta/s=\pm0.05$, and $\Delta t_{\rm start}=\pm0.3$\,fm/$c$.}
\begin{tabular}{cccccccccccccccc}
\hline\hline
$\Elab$ & \multicolumn{3}{c}{(p)S-condition}
&& \multicolumn{3}{c}{(p)M\&V-condition}
&& \multicolumn{3}{c}{$(\pi)$S-condition}
&& \multicolumn{3}{c}{$(\pi)$M\&V-condition}
 \\
\cline{2-16}
$[\agev]$
&  $t_{\rm start}$\,[fm/$c$] &$\Tfrz$\,[MeV] & $\eta/s$
&& $t_{\rm start}$\,[fm/$c$] & $\Tfrz$\,[MeV] & $\eta/s$
&& $t_{\rm start}$\,[fm/$c$] & $\Tfrz$\,[MeV] & $\eta/s$
&& $t_{\rm start}$\,[fm/$c$] & $\Tfrz$\,[MeV] & $\eta/s$\\
\hline\hline
6    & 8.41   & 80  & 0.1 && 8.41 & 80  & 0.1 && 8.93 & 110  & 0.2  && 8.93 & 110 & 0.2 \\
\hline
10.7 & 6.97   & 90  & 0.15&& 9.44 & 100 & 0.3 && 5.46 & 100  & 0.5  && 8.49 & 100 & 0.3 \\
\hline
40   & 6.22   & 130 & 0.2 && 5.94 & 140 & 0.2 && 3.93 & 150  & 0.3  && 3.93 & 150 & 0.3  \\
\hline
80   & 5.73   & 130 & 0.2 && 6.21 & 140 & 0.2 && ---  & ---  & ---  &&  --- & ---  & ---  \\
\hline
158  & 4.79   & 140 & 0.2 && 4.79 & 140 & 0.2 && ---  & ---  & ---  &&  --- & ---  & ---   \\
\hline\hline
\end{tabular}
\label{tab:fit-parameters}
\end{table*}

After considering the properties of different $\pi^{\mu\nu}$ constraints, we can try to fit the rapidity distributions and transverse momentum spectra in a wide range of bombarding energies reachable at the AGS and SPS facilities.
For each energy we vary independently the parameters $t_{\rm start}$, $\Tfrz$, and $\eta/s$. We also try various constraints on the shear stress tensors evaluating the quantity $q$ in Eq.~(\ref{piconstrain}), keeping there $C=1$, according to the strict S-condition~(\ref{q-def-S}) and the weaker M-condition (\ref{pimusic}). Recall the results for the V-conditions are identical to those for the M-condition.

As we mentioned in previous Sections~\ref{ssec:Tfo} and \ref{VHLLEMUSIC} there is a tension in the description of proton and pion rapidity distributions. Therefore, we try two strategies for data fitting.

Applying the first one, we will require the best description of the proton rapidity distribution, then tune the width of the pion rapidity distribution and proton transverse momentum spectrum.
The second strategy is to insist on the best description of the pion rapidity distribution, then tune the width of proton rapidity distribution and pion $m_T$ spectra.

\begin{figure*}
\centering
\includegraphics[width=17cm]{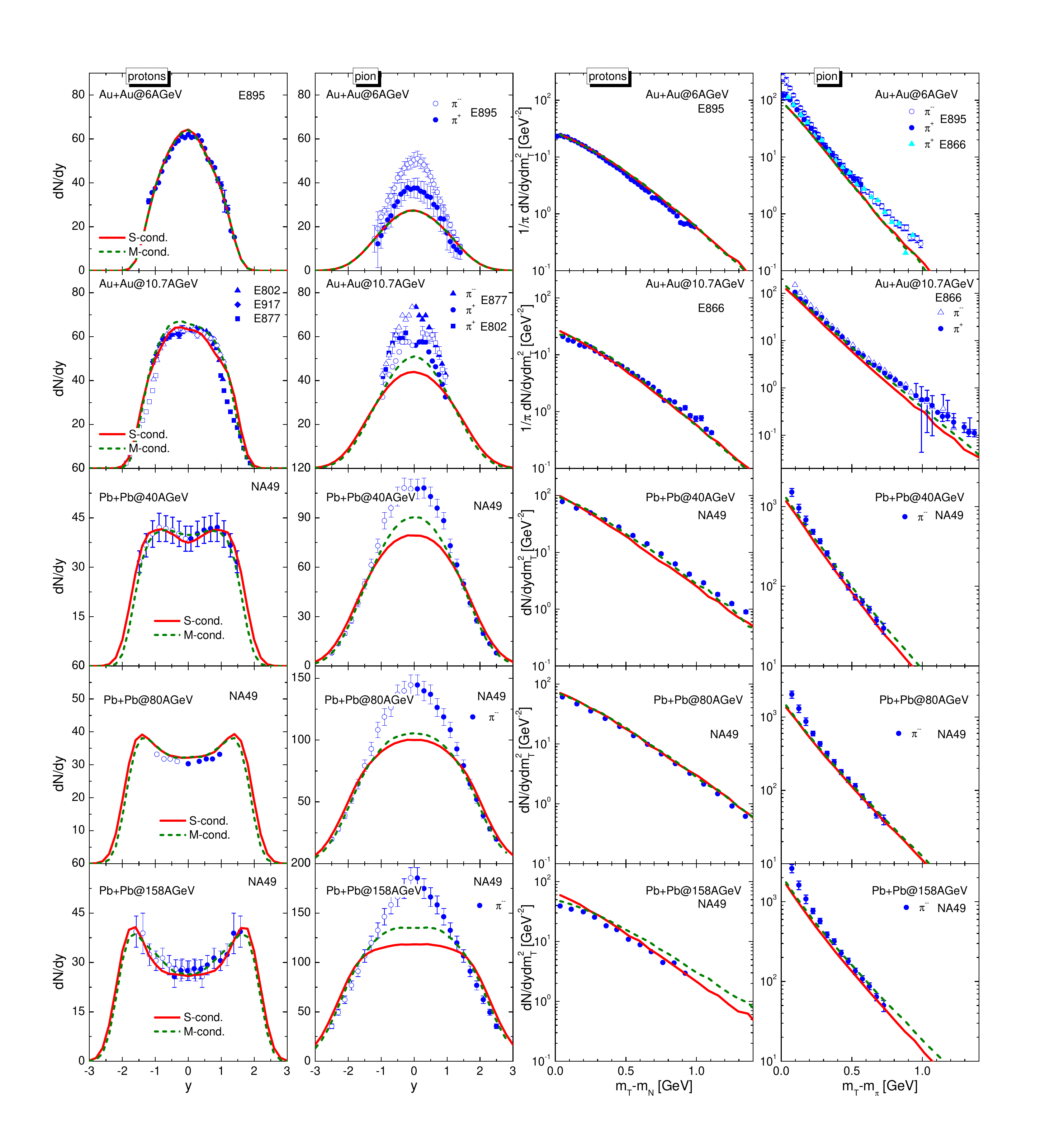}
\caption{Best fits of rapidity distributions and mid-rapidity transverse momentum spectra of protons and pions obtained by varying the values of parameters $t_{\rm start}$, $\eta/s$, and $\Tfrz$. The used fitting strategy suggests the best description of proton spectra first. The best values of fitting parameters are given in Table~\ref{tab:fit-parameters} in columns `(p)S' and  `(p)M\&V'. Experimental data are from Refs.~\cite{
SPSN1,SPSN2,SPSpiK,E895-prot,E895-pion,PRC57,PRC62,PRL86,PRC59,NPA610, PRC73, JPhysG35, NPA715}. The results of calculations done with the V-condition are identical to those for the M-condition.}
\label{fig:best-fits}
\end{figure*}

The results of the first fit strategy are presented in Fig.~\ref{fig:best-fits}.
The obtained best values of the varied parameters are collected in Table~\ref{tab:fit-parameters} in the columns indicated as (p)S- and (p)M-conditions.

Consider, first, the results obtained with the S-condition shown in Fig.~\ref{fig:best-fits} by solid lines. As we have already seen in the previous sections, we cannot simultaneously reproduce pion and proton distributions in this case.
For all considered energies we can well reproduce $\rmd N_p/\rmd y$ and $\rmd^2 N_p/\rmd y \rmd m_T$ for protons, excluding the $m_T$-spectrum at $\Elab=158\agev$ where the data for $m_T<0.3\,$GeV are overestimated.
At all considered energies, the experimental pion rapidity spectra are underestimated for $-1\lsim y\lsim 1$.
The slopes and magnitudes of the pion $m_T$-spectra are well reproduced for $m_T-m_\pi\sim 0.5\pm 0.2$\,GeV at SPS energies and for $m_T-m_\pi\sim 0.3\pm 0.2$\,GeV for AGS energies. However, the calculations do not reproduce low $m_T$ enhancements.
Different values of $\Tfrz$ are found for AGS energies but almost identical temperatures for considered SPS collisions.
In agreement with our previous results for ideal hydrodynamics~\cite{HYDHSD2015} and discussion in Sec.~\ref{param_depend}, to reproduce the proton transverse mass spectra, we need to take quite low freeze-out temperatures. As a result, obtained $\Tfrz$ values are significantly lower than predicted by a thermal statistical model~\cite{Andronic} but demonstrate a saturation at high energies.
We found a monotonic increase of $\eta/s$ parameter with an increase of the collision energy up to the value  $\eta/s\sim0.2$ for the SPS data.

Next, we apply the weaker  M-condition. It turns out that parameters of (p)M-fits are close to those obtained with the S-condition except those for the energy $\Elab=10.7\,\agev$. The larger sensitivity of observables to the $\eta/s$ value allows us to obtain higher pion rapidity distributions than for the S-condition. The quality of the description of the $m_T$ spectra remain generally the same with only a small improvement of pion spectra and a worsening for proton spectrum at $\Elab=158\,\agev$.
Surprisingly, we observe a non-monotonic behaviour in the dependence of the $t_{\rm start}$ parameters on the collision energy. In contrast to the S-condition fits where $t_{\rm start}$ decreases with an energy increase, the fit obtained with the M-condition leads to very large values of $t_{\rm start}$ and the viscosity parameter to reproduce the data at $\Elab=10.7\ \agev$.
This exception may be an evidence of a problem of the initial state for this energy obtained within PHSD~1.0.
Except for this outlier, (p)S- and (p)M-fits produce close or coinciding parameters for all other collision energies. We may interpret
this as a signal of a small degree of non-equilibrium at AGS-SPS energies.

Now we turn to the second strategy and require the best possible description of the pion rapidity spectra. Primarily, we will
apply the M-condition, since in this case the results are more sensitive to the viscosity parameter and one can potentially increase the height of the pion rapidity distribution up to the experimental values, see Fig.~\ref{fig:40AGeV-pi}.
However, we have to stress that it happens at cost of some increase of the number of cells where the elements of the $\pi^{\mu\nu}$ tensor exceed dramatically the components of the $T_{\rm id}^{\mu\nu}$ tensor.

\begin{figure}
\centering
\includegraphics[width=8.8cm]{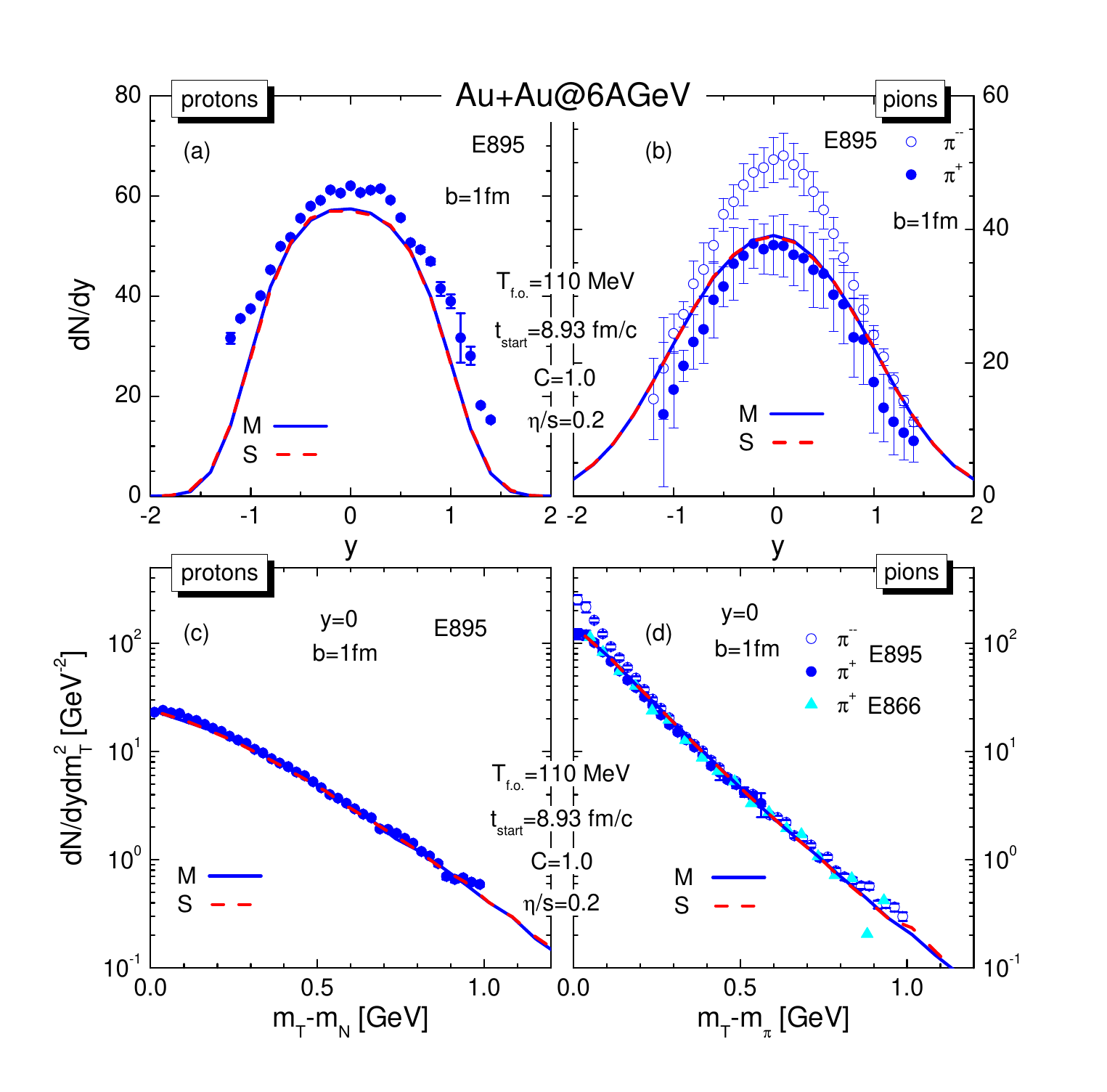}
\caption{Best fits of rapidity distributions and mid-rapidity transverse momentum spectra of protons and pions
for Au+Au collisions at $6\,\agev$ obtained by varying the values of parameters, $t_{\rm start}$, $\eta/s$, and $\Tfrz$ with the M-condition (solid line). The used fitting strategy suggests the best description of pion spectra first.
The best values of fitting parameters are given in Table~\ref{tab:fit-parameters} in column `($\pi$)M\&V'.
Dashed lines are recalculated for the S-condition with the same parameters. Experimental data are from Refs.~\cite{E895-prot,E895-pion}. Fits for the V-condition are the same as for the M-condition and are not shown.}
\label{fig:pifit-Au6AGeV}
\end{figure}

In Fig.~\ref{fig:pifit-Au6AGeV} we show the result for Au+Au collisions at $\Elab=6\,\agev$. We can find a parameter set, which describes adequately rapidity and $m_t$ distribution of $\pi^+$ mesons. The results for $\pi^-$ mesons remain underestimated especially at mid-rapidity and low $m_T$ values. The proton $m_T$ distribution is nicely reproduced but the rapidity distribution is slightly broader than the experimental data and it is not high enough. The obtained freeze-out temperature is by 30\,MeV higher than that for the proton fit strategy, and the $\eta/s$ is twice as large, see Table~\ref{tab:fit-parameters}, column '($\pi$)M'.
The transition time to the hydrodynamics, $t_{\rm start}$, is also increased by $\sim 6\%$. If we keep the same values of parameters and let the code run with the S-condition we obtain the momentum distributions shown in Fig.~\ref{fig:pifit-Au6AGeV} by dashed lines. We see that they coincide with the result for the M-condition.

\begin{figure}
\centering
\includegraphics[width=8.8cm]{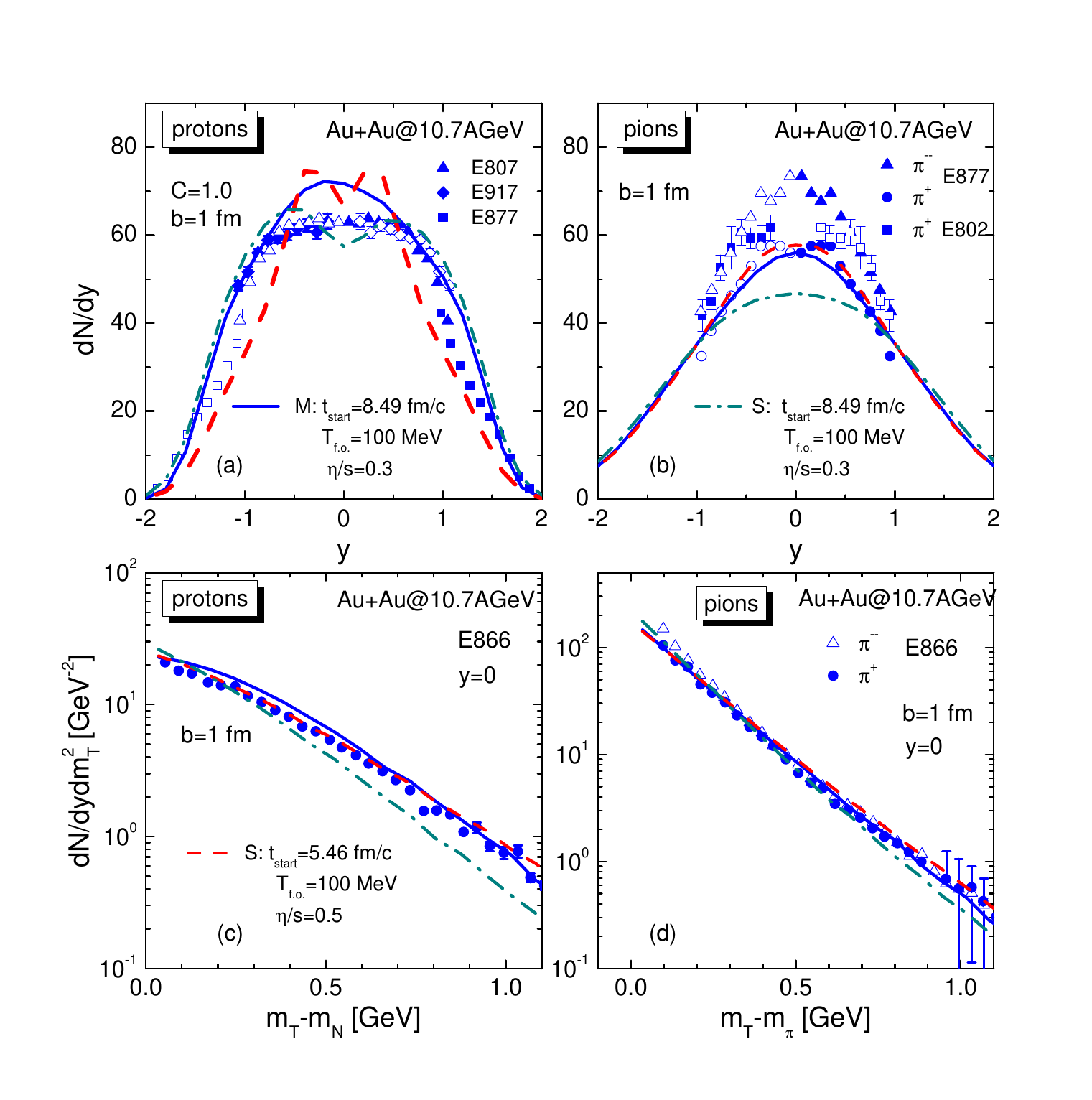}
\caption{ The same as in Fig. \ref{fig:pifit-Au6AGeV} but for $\Elab=10.7\,\agev$.
Results for the M-constraint are given by solid lines.
Dash-dotted lines are recalculated for the S-condition with the same parameters.
Dashed lines correspond to the S-condition and fine tuned parameters given in Table~\ref{tab:fit-parameters} in column `($\pi$)S'. Experimental data are from Refs.~\cite{PRC57,PRC62,PRL86}.
}
\label{fig:pifit-Au10AGeV}
\end{figure}

The result for Au+Au collisions at $\Elab=10.7\,\agev$ are shown in Fig.~\ref{fig:pifit-Au10AGeV}.
By solid lines, we plot the results obtained with the M-condition. With the increased viscosity parameter, $\eta/s=0.3$, and freeze-out temperature, $\Tfrz=100$, the rapidity and $m_T$ distributions of $\pi^+$ mesons are reasonably well reproduced. On the other hand, proton rapidity distribution is overestimated at mid-rapidity by $\sim 17\%$, and the proton $m_T$ spectra are slightly overestimated too.
If now we make a run with the S-condition and the same set of parameters, we obtain the results shown by dot-dashed lines. The maximum of the pion rapidity spectrum is reduced now by $\sim 18\%$. The quality of the description of the pion $m_T$ spectrum remains the same, but for protons, the $m_T$ spectrum falls below the experimental data points. The proton rapidity distribution decreases and comes close to experimental points. However, it develops a double-hump structure with a dip at the mid-rapidity falling below experimental points.
Now we can try to fit the pion rapidity spectrum in calculations with the S-constraint. The results are shown by dashed lines. We managed to increase the height of the $y$ spectrum up to the result obtained with the M-condition taking larger viscosity parameter $\eta/s=0.5$ (for the same $\Tfrz=100$\,MeV) and much earlier transition time $t_{\rm start}$. The proton and pion $m_T$ spectra are also well reproduced. But the rapidity distribution of proton changes dramatically it becomes higher and narrower and exhibits a double-hump structure not seen in the experiment.

\begin{figure}
\centering
\includegraphics[width=8.8cm]{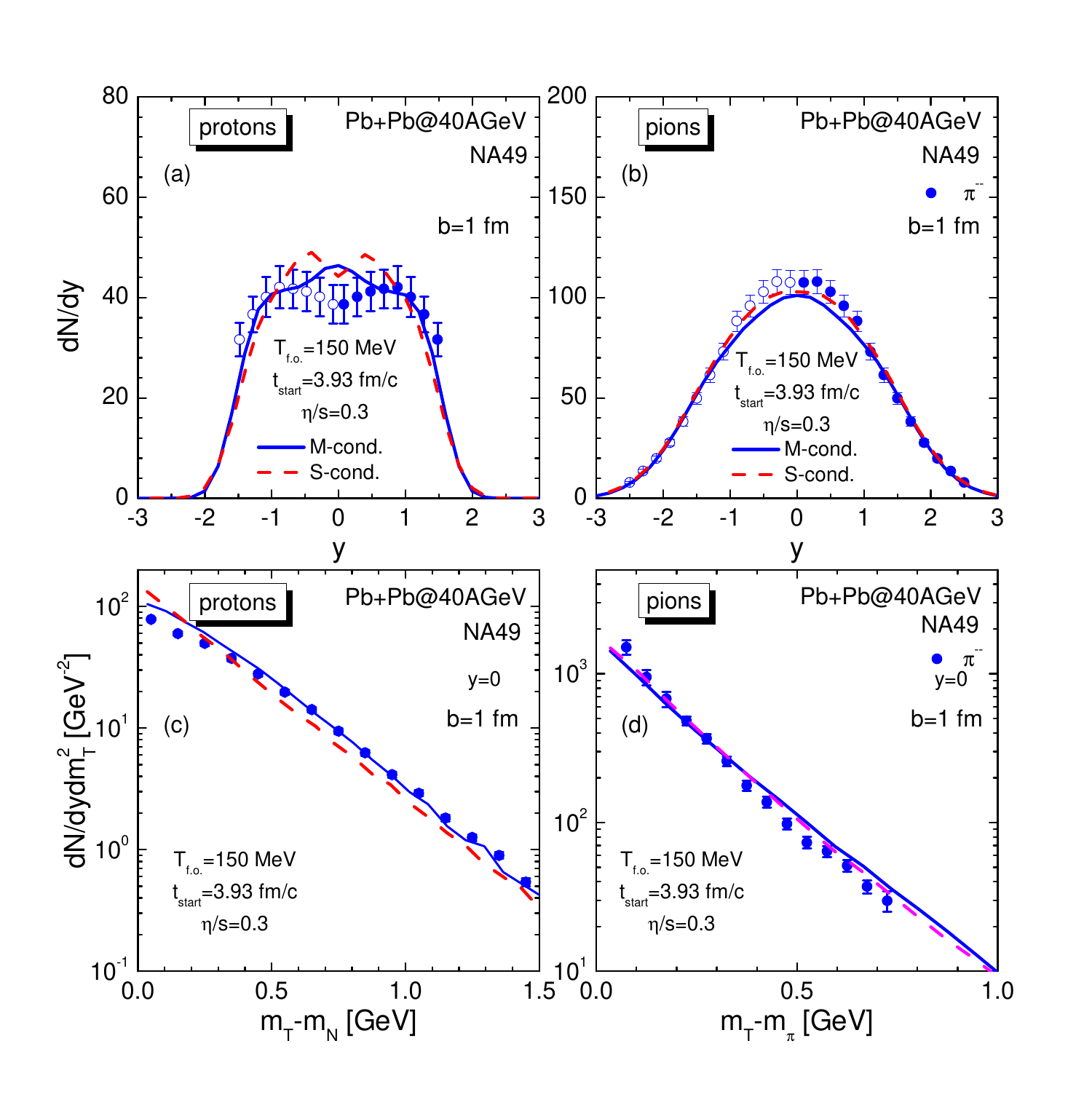}
\caption{ The same as in Figs. \ref{fig:pifit-Au6AGeV}, \ref{fig:pifit-Au10AGeV} but for $\Elab=40\,\agev$. Results for the M-constraint are given by solid lines. Dashed lines are recalculated for the S-condition with the same parameters. Experimental points are taken from Refs.~\cite{SPSN1,SPSN2,SPSpiK}.
}
\label{fig:pifit-Pb40AGeV}
\end{figure}

In Fig.~\ref{fig:pifit-Pb40AGeV} we present fits for Pb+Pb collisions at $\Elab=40\,\agev$ obtain following the second strategy when priority is given to the pion rapidity and $m_T$ distributions. Solid lines correspond to fits with the M-conditions. An increase of $\Tfrz$ up to 150\,MeV and $\eta/s=0.3$ allows the pion rapidity spectrum to pass close to the lower bound of experimental points. The transition time $t_{\rm start}=3.93$\,fm/$c $, has to be taken considerably shorter than in the case of the first (proton oriented) strategy.  However, the slope of the pion $m_T$ spectrum is not steep enough. For protons, the $m_T$-spectrum is well reproduced, but the rapidity distribution shows a bump at mid rapidity, whereas the data have a dip. Although, the experimental width of the distribution is reproduced. Dashed lines depict the result obtained with the same parameters but the S-condition. We see that the pion spectra do not change much but the proton $m_T$ spectrum goes under the experimental point for $m_T-m_N>0.3$\,GeV and above them for smaller $m_T$. As the result, the proton rapidity spectrum overestimates the data for $|y|<0.8$ and has a structure with two narrow humps.

For higher collision energies $\Elab=80\,\agev$ and $158\,\agev$ the fitting of the data following the pion-oriented strategy gives approximately the same quality of description as shown in Fig.~\ref{fig:pifit-Pb40AGeV}, however, we have found difficult to pin down a unique set of the parameter values giving the best description of experimental data. In general, using the M-condition one can tune the height of the pion rapidity spectrum by a sufficient increase of the viscosity parameter and the freeze-out temperature.
The pion $m_T$ spectra could be also reproduced except for a shortage in soft pion for $m_T-m_\pi\lsim 0.3$\,GeV There remain however severe problems with the proton rapidity distributions, which cannot be described.

Summarising the discussion of Figs.~\ref{fig:best-fits}-\ref{fig:pifit-Pb40AGeV}, we conclude that neither strategy (proton-oriented or pion oriented) allows for simultaneous description or proton and pion rapidity distribution, and the pion distribution can be described only with the weaker M-condition (or equivalently with the V-condition) when the code remains sensitive to a viscosity increase. Notice that a similar problem is encountered also in vHLLE model, see Ref.~\cite{KHPB} where a nice fit of transverse momentum spectra of protons, pions, and kaons is accompanied by an underestimation of the height of the pion rapidity distributions for SPS energies.

In the discussed AGS-SPS energy range, the detailed comparison of experimental data with different viscous-hydro approaches was made only in a couple of papers. The great success was reached within the three-fluid dynamics (3FD) model~\cite{IRT06} applied to energies $E_{\rm lab}\lsim 158\,\agev$. The 3FD approximation is a minimal way to simulate the early-stage non-equilibrium in colliding nuclei. In contrast to the conventional 1-fluid hydrodynamics, the 3FD approach takes into account a finite stopping power in a counterstreaming regime of leading baryon-rich matter at an early stage of a collision, which allows one to use a constant $t_{\rm start}$ parameter independently of $\sqrt{s}$. Formally, the model has only one free parameter for a hadronic EoS which is the formation time of the fireball. But if one uses an EoS containing a deconfinement phase transition, there appears another tuning parameter, namely, the friction of the quark phase. Comparison of the results of the 3FD model for different EoS shows that the best agreement with experimental data can be found for the case of an EoS with a smooth crossover phase transition to the quark-gluon phase. The beam-energy dependence of rapidity (not pseudorapidity !) proton spectra was found in~\cite{Iv16} to be in a good agreement with experiment at $E_{\rm lab}\lsim 10\,\agev$ for all EOS, but a mixed phase with the smooth crossover dominates definitely at higher energies. A similar situation occurred for the transverse mass spectra at the middle rapidity~\cite{Iv14}. Effects of the EOS are getting visible in more delicate characteristics, say, the energy dependence of the slopes of transverse mass spectra for identified hadrons.

The collective behavior of the nuclear fireball can also be studied using the hydrodynamics-inspired phenomenological model called the blast wave model~\cite{FB04}. The main underlying assumption of this model is that the particles in the system produced in the collisions are locally thermalized and the system expands collectively with a common radial velocity field undergoing an
instantaneous common freeze-out. While the spherically expanding source may be expected to mimic the fireball created at low energies, at higher energies a stronger longitudinal flow might lead to cylindrical geometry. For the latter case, an appropriate formalism was first developed in Ref.~\cite{SSH93}. Using a simple functional form for the phase space density at kinetic freeze-out, the authors approximated the hydrodynamical results with the boost-invariant longitudinal flow.  The common assumption for all variants of the blast
wave model is the underlying boost-invariant longitudinal dynamics. Although it is a reasonable assumption at RHIC and LHC energies, longitudinal boost-invariance does not hold well at AGS-SPS energies. Recently, a non-boost-invariant blast wave model has been developed~\cite{RBJR18}.
The model was applied in the AGS-SPS energy range to fit the rapidity distributions and transverse momentum spectra with only two parameters: the kinetic freeze-out temperature $T_{\rm f.o.}$ and the radial flow strength $\beta_T$. Authors admits that the blast-wave model cannot describe simultaneously experimental rapidity distributions and transverse momentum spectra using the same $T_{\rm f.o.}$. It is a reason to further developing of hydrodynamical (hybrid) models for the AGS-SPS-NICA-FAIR energy ranges.

\section{Conclusions}

In this work, we developed the extended version of the HydHSD (Hybrid Hadron String Dynamics) model developed in~\cite{HYDHSD2015}, which includes the effects of a shear viscosity within the Israel-Stewart hydrodynamics. Using the updated version of the hybrid model, we considered proton and pion rapidity distributions and transverse momentum spectra for $6\,\agev\leq E_{\rm lab}\leq 160\,\agev$.
As in other viscous hydrodynamic calculations, genuine inaccuracy of a numerical implementation leads to an increase of the shear stress tensor $\pi^{\mu\nu}$, Eq.~(\ref{Tmunu}), that contradicts to a perturbative character of the viscous corrections to ideal hydrodynamics. Also, codes might develop numerical instabilities~\cite{Denicol18}.
To timid the problem, a regularization scheme was suggested in the literature, which assumes the rescaling of the $\pi^{\mu\nu}$ if it exceeds the ideal energy-momentum tensor, Eq.~(\ref{T-ideal}) according to some criterion. We consider several criteria used in the literature and investigate how their applications change the results of calculations.  We use the strict (S-) condition (\ref{q-def-S}) proposed in Ref.~\cite {MNR2010}, which guarantees that each element of the $\pi^{\mu\nu}$ tensor remains smaller not more than $C$ times  the corresponding element of the $T_{\rm id}^{\mu\nu}$ tensor, see Eq.~(\ref{piconstrain}) (here $C$ is the parameter, which should be smaller than one). Also, we analyzed other conditions used in the literature: the V-condition (\ref{pivHLLE}) used in the vHLLE code~\cite{KHB2013,KHPB} and the M-condition (\ref{pimusic}) used in the MUSIC and iEBE-VISHNU codes~\cite{MUSIC,VISHNU}. We found also that in all cases the results obtained for the V-conditions are similar to the results obtained with the M-condition.

Among the details of the mode described in Section~\ref{sec:model}, such as composition of the numerical scheme, the initialization procedure, and the equation of state, we shortly discussed problems of the particlization procedure and various schemes to realize viscous corrections to the Cooper-Fry formula used to simulate particle momenta distributions at freeze-out. We used the same equation of state as in Ref.~\cite{HYDHSD2015} which was developed in Ref.~\cite{SDM09} and is the purely hadronic equation of state. We purposely refrain from variations of the equation of state before the properties and performance of the code are fully understand.

In Section~\ref{param_depend} we studied the dependence of proton and pion momentum distributions on the shear viscosity, freeze-out temperature, and the constraint regularizing the viscous stress tensor. Calculations for the S-condition shown in Fig.~\ref{fig:40AGeV-eta} particularly demonstrate that the height of the pion rapidity distributions grows with an increase of the $\eta/s$ parameter. However, the sensitivity of the distributions to the viscosity gets saturated for $\eta/s\gsim 0.2$. The sensitivity can be restored if on lets the code run with $C>1$, i.e. when viscous effects are non-perturbative, see Figs.~\ref{fig:40AGeV-C} and \ref{fig:40AGeV-C-2}. With an increase of $C$, the rapidity spectra are increased in height for pions and get deformed for protons. The slopes of $m_T$ spectra decrease also.
The independent variation of the freeze-out temperature for fixed $\eta/s=0.5$ does not influence much the pion momentum distributions but broadens the proton rapidity spectrum and flattens the slope of the proton $m_T$ spectrum.

It was shown that the code is more sensitive to the viscosity if we use the weaker V- and M-conditions instead of the S-condition. Their weakness is confirmed also numerically since the similar results for rapidity distributions can be obtained with the S-condition for a quite large value of $C$, see Figs.~\ref{fig:40AGeV-C-2}, \ref{fig:40AGeV-pi}. For example, the pion rapidity distribution calculated with the M-condition, $\eta/s=0.5$ and $C=1$, see Fig.~\ref{fig:40AGeV-C-2}, is higher than the distribution calculated with the same value of $\eta/s$ but for the S-condition and $C=20$. For weaker conditions we found also that the viscous correction term in the Cooper-Fry formula (\ref{f-CF-full}), (\ref{visc-df}) gives substantial contribution (for $\eta/s=0.5$) to the formation of final momentum distributions of both pions and protons, see Fig.~\ref{fig:CF-corr}.
For such a large value of $\eta/s$ it was proven that in the case of the M- or V-condition applied, the majority of fluid cells ($\gsim 60\%$) have non-perturbative contributions from the viscous stress tensor, see Fig.~\ref{fig:q-distrib}. This goes beyond the perturbative nature of the original hydrodynamic equations. In practice, this leads to higher temperatures of fluids and consequently to a higher freeze-out volume contributing to the pion yield, see Fig.~\ref{fig:centcell}.

Using the developed code, we performed a fit of experimental pion and proton momentum distributions for all considered collision energies. The results of our attempts to reach the best possible description are demonstrated in Fig.~\ref{fig:best-fits}, where we insisted on the best possible description of the proton rapidity distributions (proton-oriented strategy) and apply the strict S-condition and the weaker M-condition. We show that it is possible to reach the satisfactory description of proton rapidity and $m_T$ spectra, and pion $m_T$ spectra, whereas the heights of the pion rapidity distributions remain below experimental data by 20--30\%. The discrepancy becomes smaller when the M-condition is used. Within such strategy it was found that $\eta/s$ as a function of the collision energy monotonically increases from $\Elab=6\agev$ up to $\Elab=40\,\agev$ and saturates for higher SPS energies.

We tried also an alternative strategy and insist on the best description of the pion rapidity distributions. The results shown in Figs.~\ref{fig:pifit-Au6AGeV}, \ref{fig:pifit-Au10AGeV}, and \ref{fig:pifit-Pb40AGeV} follows that it can be done if large values of $\eta/s$ and the freeze-out temperature are chosen and the M-condition is used. The unique determination of parameters is possible for Au+Au collisions at $\Elab=6\,\agev$ and $10.7\,\agev$, and for Pb+Pb collisions at $40\,\agev$.  The price of the satisfactory description of pion spectra within our model is a bad description of proton ones. The discrepancy increases when the S-condition is used to fit the parameters, see Fig.~\ref{fig:pifit-Au10AGeV}.

Thus, any considered condition does not allow us to reproduce simultaneously pion and proton experimental data with good accuracy. This is in line with the results obtained in Ref.~\cite{KHPB}. We should note that for the moderate beam-energy range considered in the paper, there is no systematic comparison of predictions of one-fluid hydrodynamical models with experimental data, although a good agreement with the experiment may be reached for separate observables. Our paper partially closes this gap.

To improve the description of pion rapidity distributions, we plan such modifications of the code as the inclusion of a finite width at the stage of resonance decays that would increase pion population at low $p_T$ and at mid-rapidity, the choice of a better EoS, and the account of fluctuating (event-by-event) initial conditions will allow us to solve this problem.

\vspace{3mm}
\begin{acknowledgments}
We thank E.~Bratkovskaya and W.~Cassing for providing the HSD code and consultations. We appreciate very much extensive discussions with  Iu.~Karpenko and Yu.B.~Ivanov and constructive remarks by G.~Sandukovskaya.
The work is supported by Slovak grant VEGA-1/0348/18 and by THOR the COST Action CA15213. A.S.K and E.E.K. acknowledge the support by the Plenipotentiary of the Slovak Government at JINR, Dubna. The work of A. Khvorostukhin was supported by the RFBR grant no. 18-02-40137 and the NARD project, no. 20.80009.5007.07.
\end{acknowledgments}

\appendix

\section{Numerical realization}\label{app:numerics}

In this Appendix we discuss the numerical scheme used to integrate the hydrodynamic equations (\ref{hydrobase-J}), (\ref{hydrodecomposition}), and (\ref{ISeqs}). The 10-dimensional vector $\vec{S}$ in the right-hand side of Eq.~(\ref{shastaform}) can be written as a combinations of two 5-dimensional vectors
\begin{align}
\vec{S}=\big(\vec{S}_{\rm cons},\vec{S}_\pi\big)
\label{S-decomp}
\end{align}
corresponding to the conservation equations (\ref{hydrobase-J}) and (\ref{hydrodecomposition}),
\begin{align}
\label{Scons}
    \vc{S}_{\rm cons}\!=\!\!
    \left[
    \begin{array}{l}
 0
 \\
 -\pd_t\pi^{00} - \Div(\vec{v} P) -
 (\pd_x\pi^{0x} + \pd_y\pi^{0y} + \pd_z\pi^{0z})
 \\
-\pd_t\pi^{0x} - \pd_xP - (\pd_x\pi^{xx}+\pd_y\pi^{xy}+\pd_z\pi^{xz})
\\
-\pd_t\pi^{0y} - \pd_yP - (\pd_x\pi^{yx} +\pd_y\pi^{yy} + \pd_z\pi^{yz})
\\
-\pd_t\pi^{0z} - \pd_zP - (\pd_x\pi^{zx} + \pd_y\pi^{zy}+\pd_z\pi^{zz})
    \end{array}
    \right],
\end{align}
and to Israel-Stewart relaxation equations (\ref{ISeqs}) for viscous fields
\begin{align}
\label{Spi}
\vc{S}_\pi &= \big[Q^{xy}, Q^{xz}, Q^{yz}, Q^{yy}, Q^{zz}\big]^{\rm T}
\nonumber\\
Q^{\mu\nu} &=\pi^{\mu\nu}\left({\rm div}\vc{v} -\frac{1}{\gamma\tau_\pi}\right)+\frac{\eta}{\gamma\tau_\pi}\,W^{\mu\nu}\,.
\end{align}

In Ref.~\cite{NDHMR12} it was noted that the algorithm could become more stable if the relaxation equations are solved by a simple centered second-order differences scheme for spatial gradients on the left-hand side of Eqs.~(\ref{ISeqs}). We have tested such a separation for the full $3+1$D calculations and find out that it leads to uncontrolled solutions. The same phenomenon was observed also for calculation done in the Milne coordinates in Ref.~\cite{MHHN2014}, where the authors used also the full SHASTA method for both conservation and relaxation equations. We think that such behaviour is caused by weak steadiness of the Euler method which leads to uncontrolled inaccuracy of a numerical solution of relaxation equations.

Thus we apply the SHASTA method to all ten equations included in Eq.~(\ref{shastaform}). To reach the quadratic precision in time we use Heun's method~\cite{Heun} which allows storing fewer intermediate points than the mid-point rule.
\subsection{3$+$1D implementation of the SHASTA algorithm}\label{app:shasta}
For completeness, we provide the complete set of formulas for the 3+1 implementation of the SHASTA algorithm extending expressions provided in Ref~\cite{MNR2010}.

For lattice realization of quantities $U(x,y,z,t)$ we will use notations
$U^{[n]}_{ijk}$, where index $n$ stands for temporal steps and $i,j,k$ for spatial lattice cells in $x,y$, and $z$ directions respectively.

At the first stage of the SHASTA algorithm for the subsequent $(n +1)$th time step, one calculates the so-called transport-diffused solution
\begin{align}
\widetilde U_{ijk}^{[n+1]} &= \widetilde{\vU}_{ijk}^x + \widetilde{\vU}_{ijk}^y + \widetilde{\vU}_{ijk}^z - 2\vU_{ijk}^{[n]} + \Delta t\, {S}_{ijk},\label{U-tilde-n+1}
\end{align}
\begin{widetext}
where $\vU_{ijk}^{[n]}$ is the full solution at the previous time step and auxiliary quantities  $\tilde{\vU}_{ijk}^{x,y,z}$ are defined as
\begin{align}
\widetilde{\vU}_{ijk}^x&=\frac12\left(\big[ Q_{ijk}^{x+} \big]^2
\big(\vU^{[n]}_{i+1,jk} - \vU^{[n]}_{ijk}\big)
- \big[Q_{ijk}^{x-}\big]^2
\big(\vU^{[n]}_{ijk} - \vU^{[n]}_{i-1,jk}\big) \right)
+(Q_{ijk}^{x+}+Q_{ijk}^{x-}) \, \vU^{[n]}_{ijk},
\\
\widetilde{\vU}_{ijk}^y&=\frac12\left(
\big[Q_{ijk}^{y+}\big]^2 \big( \vU^{[n]}_{i,j+1,k}- \vU^{[n]}_{ijk} \big) - \big[Q_{ijk}^{y-}\big]^2 \big( \vU^{[n]}_{ijk}- \vU^{[n]}_{i,j-1,k} \big)
\right)
+(Q_{ijk}^{y+}+Q_{ijk}^{y-}) \, \vU^{[n]}_{ijk},
\\
\widetilde{\vU}_{ijk}^z&=\frac12\left(\big[Q_{ijk}^{z+}\big]^2
\big(\vU^{[n]}_{ij,k+1} - \vU^{[n]}_{ijk}\big) - \big[Q_{ijk}^{z-}\big]^2\big(\vU^{[n]}_{ijk} - \vU^{[n]}_{ij,k-1}\big)\right)
+ (Q_{ijk}^{z+} + Q_{ijk}^{z-})\, \vU^{[n]}_{ijk}
\end{align}
with
\begin{align}
Q_{ijk}^{x\pm}&=\frac{1/2\mp \lambda\, (v_x)^{[n]}_{ijk}}
{1\pm\lambda \,\left[ (v_x)^{[n]}_{i\pm1,jk} - (v_x)^{[n]}_{ijk} \right]},\,
Q_{ijk}^{y\pm}=\frac{1/2\mp \lambda\, (v_y)^{[n]}_{ijk}}
{ 1\pm \lambda\, \left[(v_y)^{[n]}_{i,j\pm1,k} - (v_y)^{[n]}_{ijk}\right]},\,
Q_{ijk}^{z\pm}=\frac{1/2\mp \lambda\, (v_z)^{[n]}_{ijk}}
{1\pm \lambda \,\left[(v_z)^{[n]}_{ij,k\pm 1} - (v_z)^{[n]}_{ijk} \right]}.
\end{align}
The velocity components are taken here at the $n$th time step. Here, parameter $\lambda =\Delta t/\Delta x=\Delta t/\Delta y=\Delta t/\Delta z$ is the Courant number which is the same for all special directions. In the SHASTA it is restricted to values $\lambda \le 1/2$.

Further, using the transport-diffused solution one calculates an antidiffusion flux that takes into account an anomalous diffusion
\begin{align}
A^{x,y,z}_{ijk}&=\frac{1}8 A_{\rm ad}^{x,y,z}\,\tilde{\vD}_{ijk}^{x,y,z},\quad
\tilde{\vD}_{ijk}^{x}=\tilde{\vU}_{i+1,jk}^x-\tilde{\vU}_{ijk}^x,\quad \tilde{\vD}_{ijk}^{y}=\tilde{\vU}_{i,j+1,k}^y-\tilde{\vU}_{ijk}^y,\quad
\tilde{\vD}_{ijk}^{z}=\tilde{\vU}_{ij,k+1}^z-\tilde{\vU}_{ijk}^z,
\label{Ad-mask}
\end{align}
where $A_{\rm ad}^{x,y,z}$ are the antidiffusive mask coefficients. For simplicity, one takes them to be equal for all special directions and set $A_{\rm ad}=1$ as the default value.
Next, we calculate the limited antidiffusion fluxes
\begin{align}
\widetilde{A}_{ijk}^x &= \sigma_{ijk}^x \max\Big[0,\min\Big(\sigma_{ijk}^x\widetilde{\vD}_{i+1,jk}^{x},\big|A^x_{ijk}\big|,
\sigma_{ijk}^x\widetilde{\vD}_{i-1,jk}^{x}\Big)\Big],\nl
\widetilde{A}_{ijk}^y &= \sigma_{ijk}^y
\max\Big[0,\min\Big(\sigma_{ijk}^y\widetilde{\vD}_{i,j+1,k}^{y},\big|A^y_{ijk}\big|,
\sigma_{ijk}^y\widetilde{\vD}_{i,j-1,k}^{y}\Big)\Big],\qquad
\sigma_{ijk}^{x,y,z}={\rm sgn} A^{x,y,z}_{ijk}.
\\
\widetilde{A}_{ijk}^z&=\sigma_{ijk}^z
\max\Big[0,\min\Big(\sigma_{ijk}^z\widetilde{\vD}_{ij,k+1}^{z},\big|A^z_{ijk}\big|,
\sigma_{ijk}^z\widetilde{\vD}_{ij,k-1}^z\Big)\Big].\nonumber
\end{align}
The total incoming and outgoing antidiffusive fluxes in the cell are calculated as
\begin{align}
A^{\rm in}_{ijk}&=
 \max\big(0,\widetilde A^x_{i-1,jk}\big)
-\min\big(0,\widetilde A^x_{ijk}\big)
+\max\big(0,\widetilde A^y_{i,j-1,k}\big)
-\min\big(0,\widetilde A^y_{ijk}\big)
+\max\big(0,\widetilde A^z_{ij,k-1}\big)
-\min\big(0,\widetilde A^z_{ijk}\big),\\
A^{\rm out}_{ijk}&=
 \max\big(0,\widetilde A^x_{ijk}\big)
-\min\big(0,\widetilde A^x_{i-1,jk}\big)
+\max\big(0,\widetilde A^y_{ijk}\big)
-\min\big(0,\widetilde A^y_{i,j-1,k}\big)
+\max\big(0,\widetilde A^z_{ijk}\big)
-\min\big(0,\widetilde A^z_{ij,k-1}\big).
\end{align}
The maximal and minimal values of the transport-diffused solution $\vU^{[n+1]}_{ijk}$ after the antidiffusion stage are between
\begin{align}
\widetilde{\vU}^{\min}_{ijk}&=
\min\left(\widetilde{\vU}_{ij,k-1}^{[n+1]},
          \widetilde{\vU}_{i,j-1,k}^{[n+1]},
          \widetilde{\vU}_{i-1,jk}^{[n+1]},
          \widetilde{\vU}_{ijk}^{[n+1]},
          \widetilde{\vU}_{ij,k+1}^{[n+1]},
          \widetilde{\vU}_{i,j+1,k}^{[n+1]},
          \widetilde{\vU}_{i+1,jk}^{[n+1]}\right),\\
\widetilde{\vU}^{\max}_{ijk}&=
\max\left(\widetilde{\vU}_{ij,k-1}^{[n+1]},
          \widetilde{\vU}_{i,j-1,k}^{[n+1]},
          \widetilde{\vU}_{i-1,jk}^{[n+1]},
          \widetilde{\vU}_{ijk}^{[n+1]},
          \widetilde{\vU}_{ij,k+1}^{[n+1]},
          \widetilde{\vU}_{i,j+1,k}^{[n+1]},
          \widetilde{\vU}_{i+1,jk}^{[n+1]}\right).
\end{align}
This information is then used to determine the fractions of the incoming and outgoing fluxes,
\begin{align}
F^{\rm in}_{ijk}  = \frac{\widetilde{\vU}_{ijk}^{\max} - \widetilde{\vU}_{ijk}^{[n+1]}}{A^{\rm in}_{ijk}},
\qquad
F^{\rm out}_{ijk} = \frac{\widetilde{\vU}_{ijk}^{[n+1]} - \widetilde{\vU}_{ijk}^{\min}}{A^{\rm out}_{ijk}}.
\end{align}
The final anti-diffusion fluxes are calculated as
\begin{align}
\hat A^x_{ijk}&=\widetilde A^x_{ijk}
\Big[\min(1, F^{\rm in}_{i+1,jk},F^{\rm out}_{ijk})\Theta(\widetilde A^x_{ijk})
+\min(1, F^{\rm in}_{ijk},F^{\rm out}_{i+1,jk})\Theta(-\widetilde A^x_{ijk})\Big],\\
\hat A^y_{ijk}&=\widetilde A^y_{ijk}
\Big[\min(1, F^{\rm in}_{i,j+1,k},F^{\rm out}_{ijk})\Theta(\widetilde A^y_{ijk})
+\min(1, F^{\rm in}_{ijk},F^{\rm out}_{i,j+1,k})\Theta(-\widetilde A^y_{ijk})\Big],\\
\hat A^z_{ijk}&=\widetilde A^y_{ijk}
\Big[\min(1, F^{\rm in}_{ij,k+1},F^{\rm out}_{ijk})\Theta(\widetilde A^z_{ijk})
+\min(1, F^{\rm in}_{ijk},F^{\rm out}_{ij,k+1})\Theta(-\widetilde A^z_{ijk})\Big].
\end{align}
Finally, the full solution for $n+1$ time step is given by
\begin{align}
\vU_{ijk}^{[n+1]}&=\widetilde U_{ijk}^{[n+1]}
+\big(\hat A^x_{i-1,jk}-\hat A^x_{ijk}\big)
+\big(\hat A^y_{i,j-1,k}-\hat A^y_{ijk}\big)
+\big(\hat A^z_{ij,k-1}-\hat A^z_{ijk}\big).
\end{align}
\end{widetext}

\begin{figure*}
\centering
\includegraphics[width=17cm]{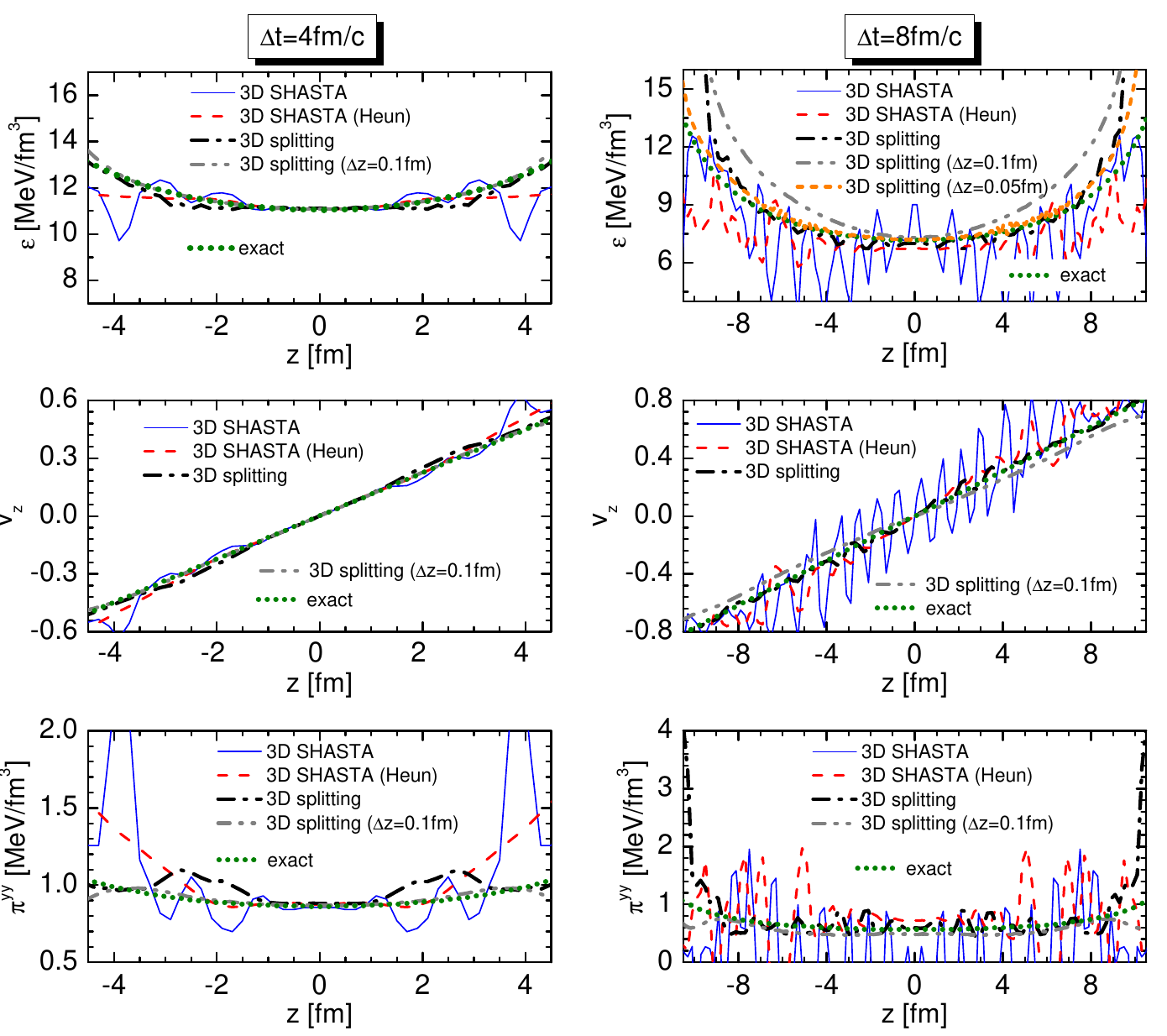}
\caption{}
\label{fig:Bjorken-num-comp}
\end{figure*}

In Fig.~\ref{fig:Bjorken-num-comp} we compare the numerical results of the 3$+$1D SHASTA code with the one-pass method in the time evolution, as given by Eq.~(\ref{U-tilde-n+1}) and of the SHASTA code improved by Heun's method with the exact results of the Bjorken model~\cite{ECHOQGP} with the viscosity $\eta=10$\,MeV/fm$^2$. The exact solutions for energy density, $\epsilon$, longitudinal velocity, $v_z$ and one component of the viscous tensor, $\pi^{yy}$ are shown in Fig.~\ref{fig:Bjorken-num-comp} by dotted lines for two times elapsed after initialization, $\Delta t=4\,{\rm fm}/c$ and $8\,{\rm fm}/c$. The solid lines show the results for the one-step 3D SHASTA. We see typical increasing oscillation expanding from the boundaries (with the square-root divergent boundary conditions) inwards the small $z$ regions. This is typical behaviour for algorithms using a single-time step approach. Applying Heun's method, we obtain much smoother behaviour of the solution shown by dashed lines. Although such scheme works well for model tasks, we observe that for actual 3$+$1D calculations in the cartesian coordinates the algorithm becomes sometimes unstable (in contrast with the 2$+$1D calculation reported in~\cite{NDHMR12}). Also the described entirely-3D approach can lead to problems with anti-diffusion as was noted in \cite{MHHN2014, NDHMR12}. To avoid the instability of the code and other possible troubles, we applied the 3D splitting method for solving the 3D problem.
\subsection{Implementation of 3D splitting SHASTA for viscous hydrodynamics}\label{app:3Dsplitshasta}
We split the 3D task (\ref{shastaform}) into three sequential 1D propagations
\begin{align}
\partial_t\vc{U}+\partial_x(v_x\vec{U})=\vec{S}^{\rm 1D}_x
\nonumber\\
\partial_t\vc{U}+\partial_y(v_y\vec{U})=\vec{S}^{\rm 1D}_y
\nonumber\\
\partial_t\vc{U}+\partial_z(v_z\vec{U})=\vec{S}^{\rm 1D}_z
\label{shasta-splitt}
\end{align}
Here, we replace the 3D propagation with three 1D propagations and the source term (\ref{S-decomp}) is split into three terms corresponding to the propagation along the separate axes. The source terms (\ref{Scons}) and (\ref{Spi}) contain time derivatives and since we replace the 3D propagation by three 1D propagations, we have to include factors $1/3$ before the corresponding terms in $\vec{S}^{\rm 1D}_{x,y,z}$. As the result, the final expression for the source term responsible for the propagation along axis $k=x,y,z$ is
\begin{align}
\label{S1Dfullsource}
\vc{S}^{1D}_k&=\Big(0,
 -\frac13\,\pd_t\pi^{00} - \pd_k (Pv_k)- \pd_k\pi^{0k},
\nl
&-\frac13\,\pd_t\pi^{0x} - \pd_x P - \pd_k\pi^{kx},
\nl
&-\frac13\,\pd_t\pi^{0y} - \pd_y P - \pd_k\pi^{ky}
\nl
&-\frac13\,\pd_t\pi^{0z} - \pd_z P - \pd_k\pi^{kz},
\nl
&\quad Q^{xy}_k, Q^{xz}_k, Q^{yz}_k, Q^{yy}_k, Q^{zz}_k
\Big)\,, 
\end{align}
where
\begin{align*}
Q^{ij}_k  =& \pi^{ij}\left(\pd_kv_k-\frac{1}{3\gamma\tau_\pi}\right)+\frac{\eta}{\gamma\tau_\pi}\,W^{ij}_k ,
\nonumber\\
W^{ij}_k  =& \frac23\,u^iu^j\,\theta_k-u^i\gD_k u^j-u^j\gD_k u^i
\nl& -\delta_{ik}\pd_k u^j-\delta_{jk}\pd_k u^i,\quad i\neq j,\\
W^{ii}_k  =& 2\Big\{\frac13\left[1+(u^i)^2\right]\,\theta_k-u^i\gD_k u^i-\delta_{ik}\pd_k u^k   \Big\},
\end{align*}
and
\begin{align}
\theta_k&\equiv\frac13\,\pd_t\gamma+\pd_ku^k,\\
\gD_k u^i&=\frac13\,\gamma\pd_t u^i+u^k\pd_k u^i.
\end{align}
As we see, after summation over index $k$
\begin{align}
\sum_k\theta_k=\theta,\,\,
\sum_k\gD_k u^i=\gD u^i,\,\,
\sum_k W^{ij}_k=W^{ij}.
\end{align}
we recover $\theta$ and $W^{\mu\nu}$ defined in Eq.~(\ref{ISeqs}), and $\gD u^i= u_\mu \pd^\mu u^i$.

Every 1D equation in Eq.~(\ref{shasta-splitt}) is solved using the standard one-dimensional SHASTA method~\cite{SHASTA,SHASTARischke}.

If in some fluid cells the relaxation time becomes smaller than the time step, $\gamma\tau_\pi<\Delta t$, then we must obtain the formal solution (\ref{pi_formal}),
\begin{align}
\pi^{ij}(t_{n+1})&=\big[\pi^{ij}(t_{n})-\eta W^{ij}\big]e^{-\Delta t/(\gamma\tau_\pi)}+\eta W^{ij}
\end{align}
after completion of the propagation along all three axes. To obtain the correct result in the 3D splitting approach, we start with the full solution at the $n$th time step, $\pi^{ij}(t_{n})$, and perform the following sequence of steps in the spatial directions staring,e.g., with the $x$ directions
\begin{align}
[\pi^{ij}(t_{n+1})]_x &= e^{-\Delta t/(\gamma\tau_\pi)} \pi^{ij}(t_{n})+\big(1-e^{-\Delta t/(\gamma\tau_\pi)}\big)\eta W^{ij}_x,
\nl
[\pi^{ij}(t_{n+1})]_y &= [\pi^{ij}(t_{n+1})]_{x}+\big(1-e^{-\Delta t/(\gamma\tau_\pi)}\big)\eta W^{ij}_y,
\nl
[\pi^{ij}(t_{n+1})]_z &= [\pi^{ij}(t_{n+1})]_{y}+\big(1-e^{-\Delta t/(\gamma\tau_\pi)}\big)\eta W^{ij}_z,
\label{pi-explit}\end{align}
so that after the third step we recover the expected expression.


When one uses 3D splitting method, it is necessary to change the order of 1D propagations to decrease numerical errors.
A similar kind of inaccuracy would appear if one uses the same relations among $\pi^{ij}$ matrix elements, e.g., $\pi^{xx}=\pi^{xx}(\pi^{yy},\pi^{zz})$ permanently. Therefore, we change the independent spatial diagonal components, $\pi^{ii}$ and $\pi^{jj}$, at every time step, see Eq.~(\ref{recover-pi}).

The results of the application of the 3D splitting scheme for the viscous 2nd-order Bjorken expansion are shown in Fig.~\ref{fig:Bjorken-num-comp} by dash-dotted lines. The calculations are performed for the same spatial step, $\Delta x=0.2$~fm, as used for the 3D~SHASTA and Heun's-improved 3D SHASTA  calculations.
We see that the numerical results are very close to the theoretical predictions and fluctuations are weaker than for the Heun's-method improved algorithm. These fluctuations decrease even further if one takes a shorter step. The results of calculations with $\Delta x=0.1$~fm are shown by dot-dot-dashed lines in Fig.~\ref{fig:Bjorken-num-comp}. For the most shown quantities, the 3D splitting results are smooth and almost coincide with the exact solutions, only the energy density $\epsilon$ for the later time, $\Delta t=8\,{\rm fm}/c$, deviates from the exact solution. This deviation vanishes if we go to a smaller step, e.g., $\Delta x=0.05$~fm, as shown by short dashes. In the actual calculations we have verified on several examples that our results do not change when we reduce the spatial steps from 0.2\,fm to 0.1\,fm.

\section{Reconstruction of local quantities and exact initialization}
The hydrodynamics code evolves the components of the energy-stress tensor and baryon current. The equation of state is formulated in the local system where the energy density and the particle number should be defined. To make a Lorentz transformation from the laboratory frame in the local rest frame one also needs to define a 4-velocity of the fluid element. If we know the components of the ideal stress tensor, $T_{\rm id}^{\mu\nu}=T^{\mu\nu}-\pi^{\mu\nu}$ and the baryon current $J^\mu=n\, u^\mu$, other quantities can be recovered as follows:
\begin{align}
n   &= J^0/\gamma\,,\quad \epsilon=T_{\rm id}^{00}- M\,v\,,
\nonumber\\
M^2 &= T^{0x}_{\rm id}\,T^{0x}_{\rm id} + T^{0y}_{\rm id}\,T^{0y}_{\rm id} + T^{0z}_{\rm id}\,T^{0z}_{\rm id}\,.
\label{recover}
\end{align}
The modulus of the fluid velocity can be found as a root of the equation
\begin{align}
v=\frac{M}{T^{00}_{\rm id}-P\big(T^{00}_{\rm id}-M\, v,J^0/\gamma\big)}
\label{recover-v}
\end{align}
and, therefore, depends on the chosen equation of state $P=P(\epsilon,n)$. The direction of the fluid velocity is determined as
\begin{align}
v^i=\frac{v}{M}\,T^{0i}_{\rm id}\,.
\label{recover-v-dir}
\end{align}

Relations (\ref{recover}), (\ref{recover-v}), and (\ref{recover-v-dir}) can be used for the 'ideal' initialization of the hydrodynamic phase
when $\pi^{\mu\nu}=0$ and $T^{0\nu}=T^{0\nu}_{\rm id}$. For the 'exact' initialization we have to solve the eigenvalue problem $u_\mu T^{\mu\nu}=\varepsilon u^\nu$,  which leads to the quartic algebraic equation
\begin{align}
\varepsilon^4 + a_1\varepsilon^3 + a_2\varepsilon^2 + a_3\varepsilon + a_4=0,
\label{e-eq}
\end{align}
where coefficients $a_{1,2,3,4}$ are functions of the energy-momentum tensor invariants,
\begin{align}
a_1&=-T^\nu_\nu=-\textrm{Tr}\,T^\mu_\nu,\quad a_2=\frac12\left(a_1^2-T^\mu_\nu T_\mu^\nu\right),
\nonumber\\
a_3&=a_1a_2-\frac{a_1^3+T^\mu_\nu T^\nu_\lambda T^\lambda_\mu}3,
\nonumber\\
a_4&=-\det T^{\mu\nu}=-\varepsilon^{\kappa\lambda\mu\nu}T^0_\kappa T^1_\lambda T^2_\mu T^3_\nu.
\label{e-eq-coeff}
\end{align}
\begin{widetext}
The corresponding velocity is calculated as
\begin{align}
v_z
&=\frac{T^{03}(\varepsilon+T^{11})(\varepsilon +T^{22})
-T^{02}T^{23}(\varepsilon+T^{11})-T^{01}T^{13}(\varepsilon +T^{22})+T^{01}T^{12}T^{23}+T^{02}T^{12}T^{13}-T^{03}(T^{12})^2}
{(\varepsilon +T^{33})(\varepsilon +T^{22})(\varepsilon+T^{11})-(T^{12})^2(\varepsilon+T^{33})
-(T^{13})^2(\varepsilon +T^{22})-(T^{23})^2(\varepsilon+T^{11})+2T^{12}T^{13}T^{23}},
\nonumber\\
v_y&=\frac{T^{02}(\varepsilon+T^{11})-T^{01}T^{12}
+\left[T^{12}T^{13}-T^{23}(\varepsilon+T^{11})\right]v_z}{(\varepsilon +T^{22})(\varepsilon+T^{11})-(T^{12})^2},
\qquad
v_x=\frac{T^{01}-T^{12}v_y-T^{13}v_z}{\varepsilon+T^{11}}.
\label{v-eq}
\end{align}
\end{widetext}
Given the four velocity, $u^\mu=\gamma(1,\vec{v})$, we calculate baryon density as
\begin{align}
n&=J^\nu u_\nu,
\label{n-eq}
\end{align}
Then knowing the equation of state $P=P(\varepsilon,n)$, we can define the ideal part of the energy-momentum tensor $T_{\rm id}^{\mu\nu}$ in Eq.~(\ref{T-ideal}). The viscous parts of the full $T^{\mu\nu}$ tensor (\ref{Tmunu}) follow as
\begin{align}
\Pi &=\frac13\,\left(\varepsilon-T^\nu_\nu\right) -P\,,
\nl
\pi^{\mu\nu}&=T^{\mu\nu}-(\varepsilon+P+\Pi)u^\mu u^\nu+(P+\Pi)g^{\mu\nu} \,,
\label{piPi-eq}
\end{align}
and the baryon diffusion current as
\begin{align}
V^\mu&=J^\mu-n u^\mu\,.
\label{V-eq}
\end{align}

In the code we use quantities $\pi^{xy}$, $\pi^{xz}$, $\pi^{yz}$, $\pi^{ii}$, and $\pi^{jj}$ as independent variables. Other components can be recovered with the help of the following expressions
\begin{align}
\pi^{0i} =& \pi^{ik} v_k+\pi^{ii} v_i+\pi^{ij} v_j,
\nonumber\\
\pi^{0j} =& \pi^{jk} v_k+\pi^{ij} v_i+\pi^{jj} v_j,
\nonumber\\
\pi^{kk} =& \frac{1}{1-v_k^2}
\Big[\pi^{0i} v_i+\pi^{0j}v_j+(\pi^{ik} v_i +\pi^{jk} v_j )v_k
\nonumber\\
&-(\pi^{ii} +\pi^{jj})
   \Big],\nonumber\\
\pi^{0k} =& \pi^{kk}v_k+\pi^{ik}v_i+\pi^{jk}v_j,
\nonumber\\
\pi^{00} =&\pi^{ii}+\pi^{jj}+\pi^{kk}\,.
   \label{recover-pi}
\end{align}
We emphasize that these expressions do not develop anomalously large values for the case of small fluid velocities.

\end{document}